\documentclass[twocolumn, astrosymb]{aastex7}

\usepackage[utf8]{inputenc}

\usepackage{natbib}
\usepackage{graphicx}
\usepackage{gensymb}
\usepackage{amsmath, amssymb}
\usepackage{mathtools}
\usepackage{xcolor}
\usepackage{hyperref}
\usepackage{physics}
\usepackage{ulem}
\usepackage{fontawesome5}
\usepackage{verbatim}

\usepackage{cellspace}
\defcitealias{GaiaCollaboration2023}{Gaia DR3}

\usepackage{multirow}

\shorttitle{GJ 105 AC}
\shortauthors{Dedrick et al.}

\newcommand{\sub}[2]{#1$_\mathrm{#2}$}

\newcommand{\msun}{M$_{\odot}~$}
\newcommand{\msune}{M$_{\odot}$}

\newcommand{\rsune}{R$_{\odot}$}

\newcommand{\teff}{\ensuremath{T_\mathrm{eff}}}

\newcommand{\atot}{a_\mathrm{tot}}

\definecolor{rvfit}{HTML}{44145C}
\definecolor{asfit}{HTML}{F9564F}
\definecolor{jfit}{HTML}{B33F62}
\definecolor{nofit}{HTML}{023047}

\usepackage{colortbl}
\newcolumntype{s}{>{\columncolor[HTML]{F4E1E6}}R}
\newcolumntype{t}{>{\columncolor[HTML]{F4E1E6}}c}

\makeatletter
\newcommand \dotfilll {\leavevmode \cleaders \hb@xt@ .44em{\hss .\hss }\hskip\z@\@plus 1filll \kern \z@}
\makeatother

\newcommand{\ry}{\textcolor{rvfit}{{\tiny{\faIcon{circle}}}}}
\newcommand{\ay}{\textcolor{asfit}{{\tiny{\faIcon{circle}}}}}
\newcommand{\jy}{\textcolor{jfit}{{\tiny{\faIcon{circle}}}}}
\newcommand{\dy}{\textcolor{jfit}{\tiny{\faIcon[regular]{circle}}}}
\newcommand{\rn}{\textcolor{rvfit}{\small{$\mathbf{\times}$}}}
\newcommand{\an}{\textcolor{asfit}{\small{$\mathbf{\times}$}}}

\providecommand{\rstar}{\ensuremath{R_\star}}

\providecommand{\teff}{\ensuremath{T_{\rm eff}}}

\providecommand{\msun}{\ensuremath{\,M_\Sun}}

\newcommand{\nstars}{805~}
\newcommand{\nsys}{401~}
\newcommand{\nbi}{398~}
\newcommand{\ntrip}{3~}
\newcommand{\neb}{353~}
\newcommand{\nast}{48~}

\begin{document}

\title{Three-dimensional Orbit and Dynamical Masses of GJ 105 AC}

\author[0000-0001-9408-8848]{Cayla M. Dedrick} 
\affiliation{Department of Astronomy and Astrophysics, The Pennsylvania State University, 525 Davey Laboratory, 251 Pollock Road, University Park, PA 16802, USA}
\affiliation{Center for Exoplanets and Habitable Worlds, The Pennsylvania State University, University Park, PA 16802, USA}
\email[show]{ckd5370@psu.edu}

\author[0000-0001-6160-5888]{Jason T. Wright}
\affiliation{Department of Astronomy and Astrophysics, The Pennsylvania State University, 525 Davey Laboratory, 251 Pollock Road, University Park, PA 16802, USA}
\affiliation{Center for Exoplanets and Habitable Worlds, The Pennsylvania State University, University Park, PA 16802, USA}
\affiliation{Penn State Extraterrestrial Intelligence Center, The Pennsylvania State University, University Park, PA 16802, USA}
\email{astrowright@gmail.com}

\author[0000-0003-3773-5142]{Jason D. Eastman}
\affiliation{Center for Astrophysics \textbar \ Harvard \& Smithsonian, 60 Garden St., Cambridge, MA 02138, USA}
\email{jason.eastman@cfa.harvard.edu}

\author[0000-0002-6096-1749]{Cullen H. Blake}
\affiliation{Department of Physics and Astronomy, University of Pennsylvania, 209 South 33rd Street, Philadelphia, PA 19104 USA}
\email{chblake@sas.upenn.edu}


\author[0000-0001-9397-4768]{Samson A. Johnson}
\affiliation{NASA JPL, 4800 Oak Grove Drive, Pasadena, CA 91109, USA}
\email{samson.a.johnson@gmail.com}

\author[0000-0002-8864-1667]{Peter Plavchan}
\affiliation{George Mason University, 4400 University Drive, Fairfax, VA 22030, USA}
\email{pplavcha@gmu.edu}

\author[0000-0002-1159-1083]{John Asher Johnson}
\affiliation{Center for Astrophysics \textbar \ Harvard \& Smithsonian, 60 Garden St., Cambridge, MA 02138, USA}
\email{jjohnson@cfa.harvard.edu}

\author[0000-0002-6228-8244]{David H. Sliski}
\affiliation{David R. Mittelman Observatory, Mayhill, NM 88339, USA}
\email{dsliski@sas.upenn.edu}

\author[0000-0003-1928-0578]{Maurice L. Wilson}
\affiliation{High Altitude Observatory, National Center for Atmospheric Research, 3080 Center Green Dr., Boulder, CO 80301, USA}
\email{maurice.wilson@cfa.harvard.edu}

\author[0000-0001-9957-9304]{Robert A. Wittenmyer}
\affiliation{Centre for Astrophysics, University of Southern Queensland, UniSQ Toowoomba, QLD 4350, Australia}
\email{Rob.Wittenmyer@unisq.edu.au}


\author[0000-0001-7139-2724]{Thomas Barclay}
\affiliation{NASA Goddard Space Flight Center, 8800 Greenbelt Road, Greenbelt, MD 20771, USA}
\email{thomas.barclay@nasa.gov}

\author[0000-0002-1160-7970]{Jonathan Horner}
\affiliation{Centre for Astrophysics, University of Southern Queensland, UniSQ Toowoomba, QLD 4350, Australia}
\email{jonathan.horner@usq.edu.au}

\author[0000-0002-7084-0529]{Stephen R. Kane}
\affiliation{Department of Earth and Planetary Sciences, University of California, Riverside, CA 92521, USA}
\email{skane@ucr.edu}

\author[0000-0002-6937-9034]{Sharon X. Wang}
\affiliation{Department of Astronomy, Tsinghua University, Beijing 100084, China}
\email{sharonw@mail.tsinghua.edu.cn}






\correspondingauthor{Cayla Dedrick}

\received{Aug. 24, 2024}
\revised{Jan. 16, 2025, Mar. 24, 2025}
\accepted{Mar. 26, 2025}
\submitjournal{ApJ}
    
\begin{abstract}
     The precision of stellar models is higher than the precision at which we are able to measure the masses of most stars, with the notable exception of binaries where we can determine dynamical masses of the component stars. In addition to well-measured stellar properties, the ideal benchmark star is far enough from its companion that its properties are indistinguishable from an otherwise identical single star. Currently, there are a handful of stars with precise ($\pm 3\%$), model-independent mass measurements that are ``effectively single" and for which we can obtain clean spectra (i.e. spectra that are not blended with a close companion). In this paper, we introduce GJ 105 AC as the newest members of this exclusive population. We present an updated orbital analysis for the long-period K3+M7 binary GJ 105 AC. We jointly analyze radial velocity (RV) and relative astrometry data, including new RVs from the Miniature Exoplanet Radial Velocity Array (MINERVA) that capture the full periapsis passage and the RV minimum of the $76.0 \pm 1.3$ year orbit for the first time. We derive precise dynamical masses of $M_1 = 0.78 \pm 0.02$ \msun and $M_2 = 0.098 \pm 0.002$ \msune. We find that of all stars with similarly precise masses ($\sim\,$2\%), GJ 105 AC stands out as having the widest on-sky separation after $\alpha$ Centauri AB, making it one of the most easily accessible to spectroscopy, as well as the the second-widest true separation, ensuring that its members are truly ``effectively single" in terms of their evolution.
\end{abstract}

\section{Introduction}\label{sec:intro}

The precision of stellar evolution models has often outpaced our ability to measure stellar properties observationally, and vice versa. Better data facilitate the improvement of models by highlighting where our theoretical understanding of stellar physics is incomplete or inaccurate. When models overtake data, we are challenged to improve our measurements.  Today, stellar structure and evolution models in many areas are ahead of the data, motivating more precise and accurate measurements of stars' ages, masses, composition, and radii.

In the case of stellar masses, there are very few stars with precise enough masses that they can be used to constrain models. An uncertainty of as little as 5\% in initial mass can make it impossible to differentiate between stellar evolution models when modeling individual stars \citep{Torres2010}. One of the only ways to measure precise, model-independent stellar masses is to measure the orbits of binary systems and compute the dynamical masses of the components. The majority of all dynamical mass measurements come from eclipsing binaries (EBs). To that end, \citet{Andersen1991} and \citet{Torres2010} previously compiled and summarized lists of well-studied detached EBs (DEBs), which they defined as those with masses known to 3\% or better. Currently, the physical properties of all DEBs that meet this criteria are tracked and kept up to date in a catalog called DEBCat\footnote{\url{https://www.astro.keele.ac.uk/jkt/debcat/}} \citep{Southworth2015}, which consists of \neb systems as of 2025 March. Additionally, we compiled a list of \nast systems with masses from astrometry \citep{Gatewood2003, Torres2010, Benedict:2016, Helminiak2019, Torres2024a, Torres2024c}, for a total of \nstars stars (\nbi binaries and \ntrip triple systems).

\begin{figure*}[htb]
    \centering
    \includegraphics{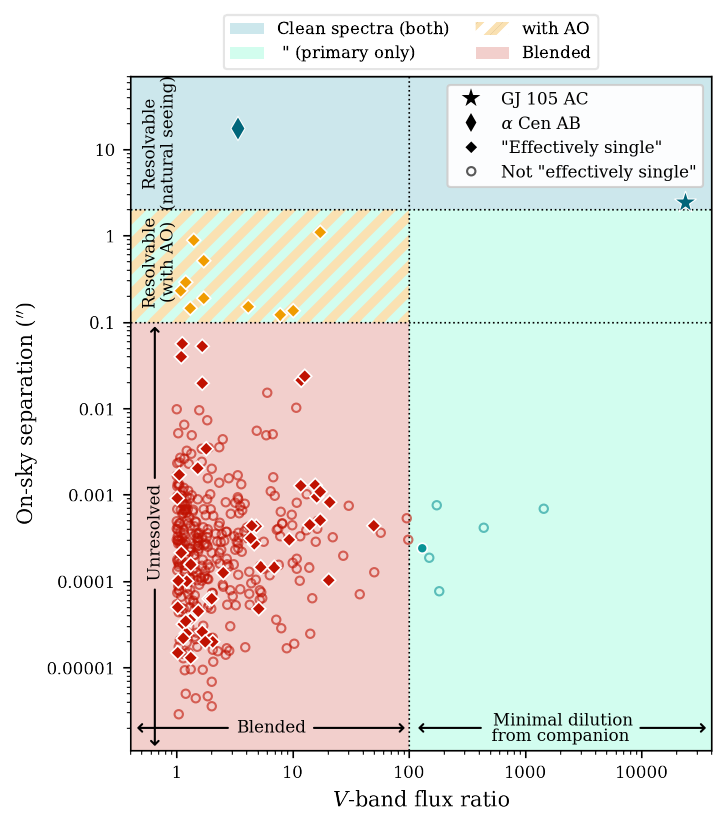}
    \label{fig:blend}
\end{figure*}

Additionally, the most useful benchmark stars are those for which we are able to obtain clean, unblended spectra, which allow us to measure the stars' projected rotation and atmospheric composition precisely. This means that there is a preference for binaries with wide separations to individually resolve both components, or large contrast ratios to obtain spectra of the primary that are effectively undiluted by the companion (see Figure \ref{fig:blend}).

Recent developments in asteroseismology and gyrochronology have begun to allow us to more accurately determine the ages of stars, which was previously very difficult, especially in the case of late-type stars. Both methods benefit from knowing that the underlying models are accurate, which is established using benchmark stars. Where possible, asteroseismology can give better age constraints for stars where the mass and radius are already well constrained. 

Precise stellar properties are crucial for constraining target selection in the context of potentially habitable exoplanets \citep{Kane2014}. Furthermore, accurate knowledge of stellar mass and age are also important for understanding the \textit{habitable history} of any planets in a system \citep{Tuchow2020}; mass and age are a probe into the location of the habitable zone as a function of time. This is important for determining how long a planet has been in the habitable zone, with implications for our expectations of its surface and atmospheric properties and the possibility of detecting biosignatures.

In this paper, we jointly model radial velocities (RVs) and relative astrometry to find a complete 3D orbital solution for the binary star GJ 105 AC to obtain new mass measurements for both components in the system, with a mass precision of $\sim\!2\%$. 


Section \ref{sec:sys} is a brief overview of GJ 105 AC and previous studies of the system's stellar parameters and orbital properties, as well as updated stellar parameters from spectral energy distribution (SED) modeling with \texttt{EXOFASTv2}. In Section \ref{sec:data}, we introduce the data sets that are used in the fit, including the new RV measurements from the Miniature Exoplanet Radial Velocity Array (MINERVA) and previously unpublished relative astrometry from Keck. The model-fitting methods are described in Section \ref{sec:methods} and the results of these fits are presented in Section \ref{sec:res}. We discuss the importance of this system as a benchmark and propose avenues for further follow-up studies in Section \ref{sec:disc}. We conclude in Section \ref{sec:conclusion}.

\section{System overview} \label{sec:sys}

GJ 105 A (HD 16160) is a bright K3V spectral standard star \citep{Keenan1989} at a distance of $7.23\,$pc \citep[$\varpi =138.34 \pm 0.32$][hereafter Gaia DR3]{GaiaCollaboration2023}, located in the thin disk \citep{Mishenina2015}. Its close companion, GJ 105 C (HD 16160 B), is an M7V star at a separation of about $3''$ from the primary \citep{Golimowski2000}. The inner binary also has a distant tertiary companion, GJ 105 B (BX Ceti), an M3.5V BY Draconis variable star \citep{Samus2009} located $165''$ ($\sim\! 1200\,$AU) to the southeast of the AC component \citep{vanMaanen1938}. The lettering convention for this triple system varies from catalog to catalog and paper to paper, which is a potential source of confusion. For clarity, we outline the conventions used by various sources in Table \ref{tab:letters}.

\begin{deluxetable*}{ccccclc}[ht]
    \tabletypesize{\footnotesize}
    \tablecaption{The lettering conventions from different catalogs for the three stars in the triple system, GJ 105.} \label{tab:letters}
    \tablehead{
        \colhead{}   & 
        \multicolumn{2}{c}{Identifier} &
        \colhead{} &
        \multicolumn{3}{c}{Multiplicity Catalogs} \\
        \cline{2-3}
        \cline{5-7}
        \colhead{\#\tablenotemark{\scriptsize{\rm *}}~~~}  & 
        \colhead{This work} &
        \colhead{SIMBAD} &
        \colhead{} &
        \colhead{WDSC} &
        \colhead{MSC} & 
         \colhead{Updated MSC}
    }
    \startdata
    1~~~ & GJ 105 A & HD 16160 & & WDS 02361+0653 A & PMSC 026307+0626 Aa & PMSC 02361+0653 A\\
    2~~~ & GJ 105 B & BX Ceti & & WDS 02361+0653 C & PMSC 026307+0626 B & PMSC 02361+0653 C\\
    3~~~ & GJ 105 C & HD 16160B & & WDS 02361+0653 B & PMSC 026307+0626 Ab & PMSC 02361+0653 B
    \enddata
    \tablerefs{Washington Double Star Catalog \citep{Mason2019}; Multiple Star Catalog \citep{Tokovinin1997};  Updated Multiple Star Catalog \citep{Tokovinin2018}}
    \tablenotetext{\rm *}{1 = Primary; 2 = Outer companion to the inner binary ($\sim\!165''$ separation); 3 = Inner companion at $\sim\!2''$ separation and the companion to which this work commonly refers}\vspace{-5pt}
\end{deluxetable*}


\subsection{Previous Orbital Analyses}

As one of the Sun's nearest neighbors, GJ 105 has been the subject of extensive observations for over 100 yr. It is only in the past four decades that the measurements of the RV and astrometric motion have been precise and reliable enough to be used to accurately determine the stars' dynamical masses. Prior to the discovery of GJ 105 C, long-term plate measurements of the absolute astrometric motion of GJ 105 A from Sproul Observatory \citep{Lippincott1973, Heintz1994} and McCormick Observatory \citep{Martin1975, Ianna1992} had indicated that the star had an unseen companion with an orbital period near 60 yr. 

Unfortunately, \citet{Lippincott1973} and \citet{Martin1975} only report the values and residuals for the orbital component of their astrometry, with the parallax and proper motion already removed. The $>70$ years of data analyzed in the two more recent absolute astrometry papers \citep{Ianna1992, Heintz1994} are not published in any form. As a result, we are unable to include any absolute astrometry data in our model. 
 
\citet{Golimowski1995} initially discovered the low-mass companion GJ 105 C at a separation of $3\farcs27$ from the primary star using ground-based adaptive optics (AO) imaging. After further follow-up observations using the Hubble Space Telescope (HST), they concluded that this companion was the most likely source of the previously observed perturbations \citep{Golimowski2000}. \citet{Roberts2018} report a very different orbital period 201 yr from fitting the same relative astrometry data used in this paper, with the exception of the NIRC2 data. The large discrepancy arises from insufficient phase coverage in the data set that does not include data from near periapsis passage. Recently, \citet{Feng2021} were the first to derive a full orbital solution for GJ 105 AC using a combined analysis of RVs and Gaia+Hipparcos astrometry. Their analysis included much of the same RV data used herein. They found a best-fit orbital solution with a period of $76.1 \pm 1.8$ years, and measured the mass of GJ 105 C to be $102.6 \pm 9.4\,$M$_{\rm J} = 0.098 \pm 0.009$ \msune. However, their data only span a portion of the periapsis passage and do not capture the ascending node (i.e. the RV minimum of the primary).

\subsection{Stellar Parameters}
\label{sec:stars}

\citet{vanBelle2009} used interferometry to directly measure the angular diameter of GJ 105 A to be $\theta_\mathrm{LD} = 0.838 \pm 0.069\,\text{mas}$ which corresponds to a physical radius of $R_\star = 0.651 \pm 0.054\,$\rsune. More recently, \citet{Boyajian2012} measured a significantly larger interferometric diameter of $\theta_\mathrm{LD} = 1.030 \pm 0.007\,\text{mas}$, or $R_\star = 0.8005 \pm 0.0057\,$\rsune. \citet{Tayar:2022} showed that interferometric radii $\lesssim 1.25$ mas can suffer systematic errors many times their formal systematic uncertainties, which may account for the large discrepancy between these values in spite of their small error bars.

\begin{figure}
    \centering
    \includegraphics[width=\columnwidth]{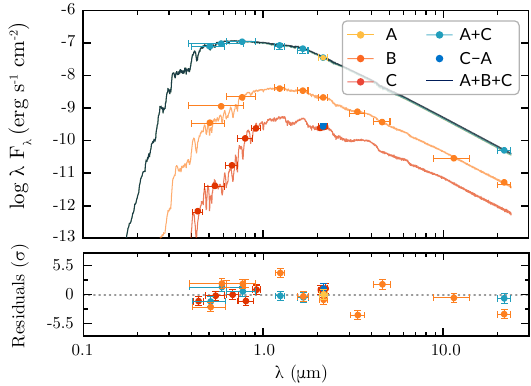}
    \caption{The SED model of the GJ 105 ABC system, showing the best fit models for each star in yellow (A), orange (B) and red (C). Blended and differential photometry are used simultaneously with the evolutionary model (see Figure \ref{fig:mistiso}) to constrain these best-fit models (see \S \ref{sec:stars}). The SED predominantly constrains the radius and temperature of each of the stars.}
    \label{fig:sed}
\end{figure}

We used a recent release of EXOFASTv2 \citep{Eastman2019}, which allows us to model an arbitrary number of stars and to simultaneously model the three stars in the system. We used the blended, resolved, and differential photometry listed in Table \ref{tab:sed} with a new SED model based on NEXTGEN stellar atmospheres \citep{Allard:2012}, and detailed transmission curves for each filter from the Spanish Virtual Observatory (SVO) \citep{Rodrigo2012,Rodrigo2020} to determine the SED model for each star shown in Figure \ref{fig:sed}. In many cases, we have rounded the photometric uncertainty up to 0.02 mag to account for systematic errors in the zero-points, stellar models, and transmission curves. 
We also opted not to include available NICMOS2 photometry on GJ 105 A, WFPC2 F336W photometry of GJ 105 C, and WISE3 photometry of A+C due to dramatic disagreements with the rest of the photometry.

\begin{figure}
    \centering
    \includegraphics{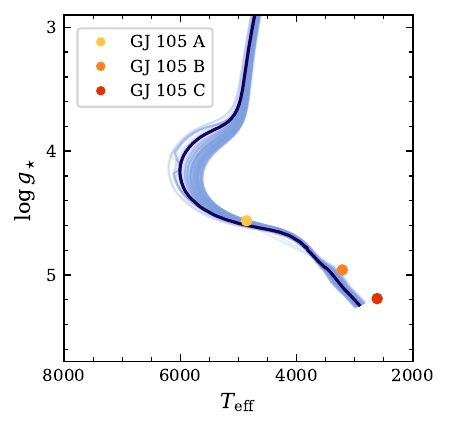}
    \caption{The MIST evolutionary model isochrone (dark blue line) of the GJ 105 ABC system at the best-fit model age of 11 Gyr and initial metallicity of -0.051 dex. The A (yellow), B (orange) and C (red) stars are overplotted on the isochrone. Given the model uncertainties of theoretical stellar evolution, they are not required to fall directly on the isochrone, but constrained with a mass-dependent Gaussian prior that penalizes the difference between our model's \teff, \rstar, and age and their MIST-computed counterparts with an uncertainty of 3.4\%, 5.8\%, and 9.5\% for stars A, B, and C, respectively (see \citet{Eastman2019} for details).}
    \label{fig:mistiso}
\end{figure}

In addition, we fixed the extinction to be zero, because GJ 105 is in the Local Bubble, and imposed priors on [Fe/H] of $-0.13 \pm 0.08$ from TICv8.2 \citep{Paegert:2021} and parallax of $138.34 \pm 0.32$ mas from \citetalias{GaiaCollaboration2023}. We used a MIST evolutionary model for each star \citep{Dotter2016}, setting the initial metallicities, distances, and ages such that they are the same for all the three stars. Unlike the typical \texttt{EXOFASTv2} MIST mass track plot that shows all ages of the star with the best-fit modeled mass and metallicity, when multiple stars are modeled with the same age, \texttt{EXOFASTv2} generates an isochrone showing all stellar masses with the same age (11 Gyr) and initial metallicity (-0.060 dex), with all modeled stars plotted on top of it, a re-creation of which is shown in Figure \ref{fig:mistiso}. The full results for all three stars from our \texttt{EXOFASTv2} fit are shown in Table \ref{tab:st_params}.

The stellar radius of GJ 105 A that we derived is $1.3\, \sigma$ larger than the interferometric radius from \citet{vanBelle2009} and $1.2\, \sigma$ smaller than the radius from \citet{Boyajian2012}. Note that our \texttt{EXOFASTv2} model did not include priors on stellar masses derived from the astrometric fit, and so our stellar parameters are both completely independent.


We also note that our isochrone constrains the age to be $11^{+1.6}_{-2.5}$ Gyr. This is consistent with and independent of the expected maximum age of 9 Gyr for systems in the thin disk. \citet{Xiang:2017} also show an empirical correlation between age, metallicity ([Fe/H]) and $\alpha$-element to iron ratio [$\alpha$/Fe], from which we can roughly estimate that the median expected age is $7-8$ Gyr for a star like GJ 105 A, with [Fe/H] = 0.09 and [$\alpha$/Fe] = 0.14 \citepalias{GaiaCollaboration2023}. This estimate is in good agreement with ours, differing from ours by at most $1.6\, \sigma$.

\begin{deluxetable}{llcchc}[ht]
    \tabletypesize{\footnotesize}
    \tablecaption{Photometry of GJ 105 used in SED fit} \label{tab:sed}
    \tablehead{
    \colhead{Instrument} & \colhead{Bandpass} & \colhead{Magnitude} & \colhead{Error} & \nocolhead{Catalog Error} & \colhead{Star}
}
    \startdata
    \multirow{6}{*}{2MASS$^1$} & J &  4.152 & 0.040 & 0.264 & A$\,+\,$C\\
     & H &  3.657 & 0.040 & 0.244 & A$\,+\,$C\\
     & \sub{K}{s} &  3.481 & 0.040 & 0.208 & A$\,+\,$C\\
     & J &  7.333 &	0.020 & 0.018 & B\\
     & H &  6.793 &	0.038 & 0.038 & B\\
     & \sub{K}{s} &  6.574 &	0.020 & 0.020 & B\\
    \hline
    \multirow{5}{*}{WISE$^2$} & W4 & 3.383 &	0.020 & 0.020 & A$\,+\,$C\\
    & W1 &  6.481 &	0.036 & 0.036 & B\\
    & W2 &  6.116 &	0.024 & 0.024 & B\\
    & W3 &  6.141 &	0.020 & 0.015 & B\\
    & W4  & 6.040 &	0.057 & 0.047 & B\\
    \hline
    \multirow{6}{*}{Gaia DR3$^3$} & 
      G           &  5.498  & 0.020 & 0.002884 & A$\,+\,$C\\
    & \sub{G}{BP} &  6.019  & 0.020 & 0.002967 & A$\,+\,$C\\
    & \sub{G}{RP} &  4.821  & 0.020 & 0.005203 & A$\,+\,$C\\
    & G           & 10.333  & 0.020 & 0.002849 & B\\
    & \sub{G}{BP} & 11.952  & 0.020 & 0.003521 & B\\
    & \sub{G}{RP} &  9.096  & 0.020 & 0.003935 & B\\
    \hline
    \multirow{5}{*}{HST/WFPC2$^4$} & F439W & 19.17 & 0.05 & 0.05 & C\\
    & F555W &  16.77 &   0.08 & 0.08 &   C\\  
    & F675W &  14.68 &   0.08 & 0.08 &   C\\
    & F814W &  12.26 &   0.03 & 0.03 &   C\\
    & F850LP & 11.17 &   0.03 & 0.03 &   C\\
    \hline
    \multirow{2}{*}{HST/NICMOS2$^5$} & F207M & 8.96 & 0.05 & 0.05 & C\\
    & F222M &  8.65 &   0.05 & 0.05 &   C\\  
    \hline
    \multirow{2}{*}{PHARO$^6$} & \sub{K}{s} &   3.48 &    0.21 & 0.21 & A\\   
    & \sub{K}{s} &   8.77 &    0.22 & 0.22 & C\\
    \hline
    Lick/IRCAL$^7$ & BrG & 5.142 & 0.1 & 0.1 & C$\,-\,$A
\enddata
\tablerefs{1 - \citet{Skrutskie:2006}, 2 - \citet{Cutri:2013}, 3 - \citetalias{GaiaCollaboration2023}, 4 - \citet{Golimowski2000}, 5 - \citet{Dieterich2012}, 6 - \citet{Tanner2010}, 7 - \citet{Rodriguez2015}}
\end{deluxetable}

\begin{deluxetable*}{lcrrr}[ht]
    \tablecaption{Median values and 68\% confidence interval for GJ 105, created using \texttt{EXOFASTv2} commit number 63b9f6ff\label{tab:st_params}}
    \tabletypesize{\footnotesize}
    \tablehead{ \colhead{Parameter} &
                \colhead{Description} & 
                \multicolumn{3}{c}{Values}
                }
    \startdata
         &  & \multicolumn{1}{c}{A} & \multicolumn{1}{c}{B} & \multicolumn{1}{c}{C}\\
         \cline{3-5}
        ~~~~$M_\star$ \dotfilll & Mass (\msune) \dotfilll & $0.737^{+0.024}_{-0.023}$ & $0.277 \pm 0.024$ & $0.1123^{+0.011}_{-0.0083}$ \\
        ~~~~$R_\star$ \dotfilll & Radius (\rsune) \dotfilll & $0.730^{+0.023}_{-0.022}$ & $0.289^{+0.012}_{-0.011}$ & $0.1329^{+0.0057}_{-0.0055}$\\
        ~~~~$R_{\star, \mathrm{SED}}$ \dotfilll & Radius\tablenotemark{\scriptsize{\rm 1}} (\rsune) \dotfilll & $0.726 \pm 0.016$ & $0.2965^{+0.0045}_{-0.0043}$ & $0.1353^{+0.0026}_{-0.0030}$\\
        ~~~~$L_\star$ \dotfilll & Luminosity \dotfilll & $0.2734^{+0.0087}_{-0.0085}$ & $0.00795 \pm 0.00023$ & $0.000758 \pm 0.000033$\\
        ~~~~$F_\mathrm{Bol}$\dotfilll & Bolometric Flux (cgs) \dotfilll &$1.674^{+0.053}_{-0.052} \times 10^{-7}$&$4.87\pm0.14 \times 10^{-9}$&$4.64\pm0.20 \times 10^{-10}$\\
        ~~~~$\rho_\star$\dotfilll &Density (cgs)\dotfilll &$2.67^{+0.24}_{-0.21}$&$16.1^{+1.8}_{-1.6}$&$68.1^{+9.3}_{-8.1}$\\
        ~~~~$\log{g}$\dotfilll &Surface gravity (cgs)\dotfilll &$4.579^{+0.025}_{-0.024}$&$4.957^{+0.036}_{-0.035}$&$5.245^{+0.043}_{-0.041}$\\
        ~~~~$T_{\rm eff}$\dotfilll &Effective temperature (K)\dotfilll &$4886^{+72}_{-71}$&$3205\pm59$&$2626\pm50$\\
        ~~~~$T_{\rm eff, SED}$\dotfilll &Effective temperature$^{1}$ (K)\dotfilll &$4898^{+47}_{-45}$&$3167^{+20}_{-19}$&$2604.6^{+7.1}_{-3.4}$\\
        ~~~~$[{\rm Fe/H}]$\dotfilll &Metallicity (dex)\dotfilll &$-0.090^{+0.072}_{-0.070}$&$-0.01^{+0.11}_{-0.10}$&$0.006\pm0.068$\\
        ~~~~$[{\rm Fe/H}]_{0}$\dotfilll &Initial Metallicity$^{2}$ \dotfilll &$-0.051^{+0.071}_{-0.070}$&$-0.051^{+0.071}_{-0.070}$&$-0.051^{+0.071}_{-0.070}$\\
        ~~~~Age\dotfilll &Age (Gyr) \dotfilll &$11.0^{+1.6}_{-2.4}$&$11.0^{+1.6}_{-2.4}$&$11.0^{+1.6}_{-2.4}$\\
        ~~~~EEP\dotfilll &Equal Evolutionary Phase$^{3}$ \dotfilll &$348.7^{+4.7}_{-6.3}$&$269.3^{+6.0}_{-6.2}$&$244.9^{+3.5}_{-5.3}$\\
        ~~~~$A_V$\dotfilll &$V$-band extinction (mag)\dotfilll &$0.00$&$0.00$&$0.00$\\
        ~~~~$\sigma_{\rm SED}$\dotfilll &SED photometry error scaling \dotfilll &$2.11^{+0.42}_{-0.31}$&--&--\\
        ~~~~$\varpi$\dotfilll & Parallax (mas) \dotfilll &$138.34\pm0.32$&$138.34\pm0.32$&$138.34\pm0.32$\\
        ~~~~$d$\dotfilll &Distance (pc)\dotfilll &$7.229\pm0.017$&$7.229\pm0.017$&$7.229\pm0.017$
    \enddata
    \tablecomments{See Table 3 in \citet{Eastman2019} for a detailed description of all parameters}\vspace{-4pt}
    \tablenotetext{1}{This value ignores the systematic error and is for reference only}\vspace{-4pt}
    \tablenotetext{2}{The metallicity of the star at birth}\vspace{-4pt}
    \tablenotetext{3}{Corresponds to static points in a star's evolutionary history. See \S2 in \citet{Dotter2016}\vspace{-4pt}}
\end{deluxetable*}

\section{Data} \label{sec:data}
\subsection{Radial Velocities}\label{ssec:rvs}

 MINERVA is a robotic observatory located at Fred Lawrence Whipple Observatory (FLWO) on Mount Hopkins, Arizona. The MINERVA telescope array is composed of four 0.7 m PlaneWave CDK-700 telescopes (henceforth T1, T2, T3, and T4.) The light from the telescopes is fiber-fed to a high-resolution ($R \sim 80,000$) KiwiSpec spectrograph \citep{Barnes2012, Gibson2012} that is calibrated using an iodine cell. The traces are imaged using a 2k $\times$ 2k detector that spans 26 echelle orders and a wavelength range of 500 – 630 nm.  MINERVA is able to achieve a typical RV precision of 5-10 m/s for bright ($V<8$) targets. The full description of the telescopes, instruments, and enclosures can be found in \citet{Swift2015}, with updates in \citet{Wilson2019}. 

We introduce 569 new RV measurements of GJ 105 A from MINERVA that span from 2017 October 29 to 2022 January 13. These data are included in Table \ref{tab:rv}. The exposure time of these observations was $1800\,$s. We are able to ignore any potential contamination of the spectrum from the secondary because of the very high contrast ratio ($\sim 10^4$) at these optical wavelengths between the two stars. We note that there is an RV offset on the order of tens of meters per second between the four MINERVA telescopes, and therefore we treated them each as separate instruments in the RV and joint fits, with their own RV zero-point and instrumental jitter parameters. We chose not to include the data from T4 because the RV baseline of these data appears to shift between seasons.

The MINERVA data are especially significant in constraining the period of this system because it is in these data that we observe the RV minimum and the full periapsis passage for the first time. It is difficult to constrain the orbital parameters of long-period high-eccentricity systems when the RV data set only covers a relatively flat portion of the orbit \citep{Wright2007}.

We also use the following RV data from the literature: 62 RVs from the Hamilton spectrograph at Lick Observatory taken between 1987 September 10 and 2011 October 12 \citep{Fischer2014}, 43 RVs from the High Accuracy Radial Velocity Planet Searcher (HARPS) taken between 2003 October 27 and 200 September 6 \citep{Butler2017}, 115 RVs from the High-Resolution Echelle Spectrograph on Keck (HIRES) taken between 2006 August 14 and 2014 December 11 \citep{Santos2010}, and 159 RVs from the Automated Planet Finder (APF) between 2013 October 25 and 2018 December 14 \citep{Feng2021}.

\begin{deluxetable}{cccc}[htb]
\tabletypesize{\footnotesize}
\tablecaption{Radial velocity measurements of GJ 105 A used in this work, including new data from MINERVA.\label{tab:rv}}
\tablehead{
\\[-15pt]
  \colhead{\hspace{1em}Time} &
  \colhead{\hspace{2em}RV\hspace{2em}} &
  \colhead{\hspace{2em}$\sigma_\mathrm{RV}$} &
  \colhead{\hspace{1em}Instrument} \\[-5pt]
    \colhead{\hspace{1em}(BJD)}   & 
  \colhead{\hspace{2em}(m/s)} &
  \colhead{ \hspace{2em}(m/s)} &
  \colhead{}
}
\startdata
\hspace{1em}2447048.87 & \hspace{2em}216.62 &  \hspace{2em}11.40 &  \hspace{1em}Lick \\
\hspace{1em}2447374.95 & \hspace{2em}238.38 &  \hspace{2em}12.15 &  \hspace{1em}Lick \\
\hspace{1em}2452939.74 & \hspace{2em}344.00 &  \hspace{2em}2.03 &  \hspace{1em}HARPS \\
\hspace{1em}2452945.70 & \hspace{2em}344.56 &  \hspace{2em}2.09 &  \hspace{1em}HARPS \\
\hspace{1em}2453962.11 & \hspace{2em}339.37 &  \hspace{2em}3.22 &  \hspace{1em}HIRES \\
\hspace{1em}2453962.11 & \hspace{2em}339.69 &  \hspace{2em}3.20 &  \hspace{1em}HIRES \\
\hspace{1em}2456590.98 & \hspace{2em}118.22 &  \hspace{2em}1.63 &  \hspace{1em}APF \\
\hspace{1em}2456591.84 & \hspace{2em}116.62 &  \hspace{2em}1.42 &  \hspace{1em}APF \\
\hspace{1em}2458055.76 & \hspace{2em}-745.38 &  \hspace{2em}1.94 &  \hspace{1em}T1 \\
\hspace{1em}2458056.77 & \hspace{2em}-741.89 &  \hspace{2em}1.75 &  \hspace{1em}T1 \\
\hspace{1em}2458419.82 & \hspace{2em}-991.96 &  \hspace{2em}2.90 &  \hspace{1em}T2 \\
\hspace{1em}2458439.88 & \hspace{2em}-1012.87 &  \hspace{2em}2.62 &  \hspace{1em}T2 \\
\hspace{1em}2458402.74 & \hspace{2em}-957.03 &  \hspace{2em}2.76 &  \hspace{1em}T3 \\
\hspace{1em}2458410.89 & \hspace{2em}-943.93 &  \hspace{2em}2.39 &  \hspace{1em}T3 \\
\hspace{2em}$\vdots$  &  \hspace{2em}$\vdots$ &  \hspace{2em}$\vdots$ &  \hspace{1em}$\vdots$
\enddata
\tablecomments{Table 2 is published in its entirety in the machine-readable format. A portion is shown here for guidance regarding its form and content.}
\end{deluxetable}

\subsection{Relative Astrometry}\label{ssec:relas}
We have a total of 13 relative astrometry measurements from 27 yr of observations taken using seven different instruments (Table \ref{tab:as}). Ten of these measurements are reported in the Washington Double Star Catalog \citep[][and references therein]{Mason2001, Mason2019}. These data are from HST and high-resolution ground-based imagers.

We also include measurements based on unpublished observations made with the NIRC2 imager behind the AO system of the Keck II telescope. Using the Keck Observatory Archive, we downloaded images gathered as part of three different programs (C17N2 --- PI Kulkarni; N096N2L --- PI Barclay, N179 --- PI Gonzales). All of these images were gathered in the NIRC2 narrow mode, with a plate scale of approximately $0.01''$ pixel$^{-1}$. These images were taken in three different narrow bands, all centered near 2.2$\mu$m (N179 --- 17 images in  Brackett-$\gamma$; C17N2 --- 11 images in H2$_{\nu=2-1}$; N096N2L --- 16 images in K$_{\rm{cont}}$) and with exposure times $<0.1$~s to prevent saturation. We analyzed raw images by estimating the centroids of GJ 105 A and GJ 105 C in each individual image by fitting a Gaussian function to the marginal flux distributions in $X$ and $Y$ independently. Then, we used the WCS information contained in the header of each image to estimate the angular separation and position angle of the binary in each image and applied the astrometric distortion model described in \citet{Yelda2010}. We note that the absolute sizes of these distortion corrections over the central portion of the NIRC2 array where the binary was imaged are small, typically less than 2~mas. While the solution from \citet{Yelda2010} was determined from data taken prior to a NIRC2 realignment in 2015 and a new distortion correction for more recent observations was determined by \citet{Service2016}, we applied the older distortion solution to all of our observations since the differences between the two solutions are typically less than 1 mas in the central region of the detector. Finally, we averaged all measurements from a given epoch and estimated a statistical uncertainty based on the standard deviations of the separation and position angle estimates from the individual images within an epoch. 

For all of the NIRC2 archival measurements we assume a final error of $0.5\degree$ in position angle and 3 mas in separation that is a combination of statistical and systematic uncertainties.

\begin{deluxetable*}{lDDDDcccc}[tb]
\tabletypesize{\footnotesize}
\tablecaption{Measurements of the position of GJ 105 C relative to GJ 105 A \label{tab:as}}
\tablehead{
\\[-15pt]
  \multicolumn{11}{c}{Published Data} & \colhead{} & \colhead{Re-weighted Err.\tablenotemark{\rm \scriptsize{1}}}\\ \cline{1-11} \cline{13-13}
  \colhead{Time}   & 
  \multicolumn2c{$\theta$} &
  \multicolumn2c{$\sigma_{\theta}$} &
  \multicolumn2c{$\rho$} &
  \multicolumn2c{$\sigma_{\rho}$}  &
  \colhead{~~~~~~~~~Inst.\tablenotemark{\rm \scriptsize{2}}~~~~~~~~~}& 
  \colhead{~~~~~~~~~Ref.~~~~~~~~~~} & \colhead{} &
  \colhead{$\sigma^*_{\rho}$}\\[-5pt]
  \colhead{(BJD)}   & 
  \multicolumn2c{($\degree$)} &
  \multicolumn2c{($\degree$)} &
  \multicolumn2c{($''$)} &
  \multicolumn2c{($''$)}  &
  \colhead{} &
  \colhead{} &
  \colhead{} &
  \colhead{($''$)}
}
\decimals
\startdata
    2449288.01 & 287.0 & 2.0 & 3.30 & 0.12     & AOC \dotfilll & \citet{Golimowski1995} & & 0.079\\
    2449653.01 & 287.0 & 2.0 & 3.24 & 0.14     & AOC \dotfilll & \citet{Golimowski1995} & & 0.079\\
    2449723.50 & 289.65 & 0.26 & 3.394 & 0.01  & WFPC2 \dotfilll & \citet{Golimowski2000} & & 0.041 \\
    2450789.73 & 293.80 & 0.24 & 3.223 & 0.008 & WFPC2 \dotfilll & \citet{Golimowski2000} & & 0.041 \\
    2450818.50 & 293.97 & 0.24 & 3.221 & 0.008 & WFPC2 \dotfilll & \citet{Golimowski2000} & & 0.041\\
    2450818.50 & 293.5 & 0.4 & 3.22 & 0.036    & NICMOS \dotfilll & \citet{Dieterich2012} & & 0.039  \\
    2452895.30 & 300.1 & 1.0 & 2.73 & 0.01     & AEOS \dotfilll & \citet{Roberts2018} &  & 0.024 \\
    2453364.91 & 304.0 & 1.0 & 2.60 & 0.01     & AEOS \dotfilll & \citet{Roberts2018} &  & 0.024 \\
    2453221.00 & 306.0 & 0.2 & 2.63 & 0.12     & PHARO \dotfilll & \citet{Tanner2010} & & 0.034  \\
    2455413.37 & 318.49 & 0.7 & 1.73 & 0.02    & IRCAL \dotfilll & \citet{Rodriguez2015} & & 0.089 \\
    2453227.58 & 305.01 & 0.5 & 2.69 & 0.003   & NIRC2 \dotfilll & this work & & 0.032 \\
    2456513.61 & 340.30 & 0.5 & 1.22 & 0.003   & NIRC2 \dotfilll & this work & & 0.032 \\
    2459101.45 & 135.92 & 0.5 & 1.04 & 0.003   & NIRC2 \dotfilll & this work & & 0.032 
\enddata
    \tablenotetext{1}{See Section \ref{sec:resigma}.}
    \tablenotetext{2}{AOC = Adaptive Optics Coronagraph on the Palomar 60'', 
    WFPC2 = HST Wide Field Planetary Camera 2, NICMOS = HST Near Infrared Camera and Multi-Object Spectrograph, 
    AEOS = Advanced Electro-Optical System telescope, 
    PHARO = Palomar High Angular Resolution Observer on the Hale Telescope, 
    IRCAL = Infrared Camera for AO at Lick}
\end{deluxetable*}

\section{Model Fitting} \label{sec:methods}

\subsection{Deriving consistent equations for 3D orbits}

When modeling Keplerian orbits in three dimensions, it is important to ensure that we are being consistent in our use of the orbital angles $\omega$ and $\Omega$. There is potential for confusion here owing to the implicit left-handedness of the coordinate system that has been adopted by the exoplanet community. Because the literature is inconsistent about this convention, and to ensure reproducibility of our results and tractability of our analysis, we briefly rederive the equations used in the RV and relative astrometry models here.

In the observer coordinate frame,\footnote{$+X$ is north (+decl.), $+Y$ is east (+R.A.) and $+Z$ is away from the observer, along the line of sight.}, the position of a body in a Keplerian orbit about the barycenter of the system is given by

\begin{equation} \label{eq:xyz}
  \begin{split}
    X_\circ &= r_\circ \qty[\cos{\qty(\nu+\omega_\circ)}\cos{\Omega} - \sin{\qty(\nu+\omega_\circ)}\sin{\Omega}\cos{i}] \\
    Y_\circ &= r_\circ \qty[\cos{\qty(\nu+\omega_\circ)}\sin{\Omega} + \sin{\qty(\nu+\omega_\circ)}\cos{\Omega}\cos{i}]\\
    Z_\circ &= r_\circ \sin\qty(\nu+\omega_\circ)\sin{i}
  \end{split}
\end{equation}

\begin{equation}
    r_\circ = \frac{a_\circ\, (1 - e^2)}{1 + e \cos \nu}
\end{equation}

Here the generic subscript ``$\circ$" is used to denote which terms are specific to the binary component to which you are referring. The orbit can be described by orbital elements that describe the shape and orientation of an elliptical orbit. Parameter $\omega_\circ$ is the argument of periapsis of the body, $\Omega$ is the longitude of the ascending node, $i$ is the inclination, $a_\circ$ is the semi-major axis of the object, and $e$ is the eccentricity. 

The final variable, $\nu$, is the true anomaly. The true anomaly describes the object's angular position on the path of its orbit at a given time, and it is a function of the period $P$, time of periapsis passage $T_0$, and eccentricity $e$. Solving for $\nu$ at some time $t$ must be done numerically. We use kepler.py\footnote{\url{https://github.com/dfm/kepler.py}}, a stand-alone version of the numerical Kepler's equation solver from \textsf{exoplanet} \citep{exoplanet:joss, exoplanet:zenodo} to convert to true anomaly in our model.







The RV of the primary is positive when the star is moving away, and therefore

\begin{align}\label{eq:rv}
    \dv{Z_1(t)}{t} = K_1\, [\cos(\nu(t) + \omega_1) + e \cos\omega_1] 
\end{align}

\noindent where $\omega_1$ and is the argument of periapsis of the primary and $K_1$ is its velocity semi-amplitude, which can be written as a function of the previously described orbital elements

\begin{equation}\label{eq:k1}
    K_1 = \frac{2 \pi}{P}\frac{a_1 \sin i}{\sqrt{1 - e^2}}
\end{equation}

The relative position of the secondary with respect to the primary, from the point of view of an observer, is

\begin{equation}
    \begin{split}
        X_\mathrm{rel} &= X_2 - X_1 \\
        Y_\mathrm{rel} &= Y_2 - Y_1
    \end{split}
\end{equation}

 Bearing in mind that $\omega_1 = \omega_2 - \pi$ and the total separation $\atot = a_1 + a_2$, this simplifies to 

\begin{multline} \label{eq:xsky}
    X_\mathrm{rel}(t) = r_\mathrm{tot}\, \big[\cos(\nu(t) +\omega_2)\cos{\Omega}\\
     - \sin(\nu(t) + \omega_2)\sin{\Omega}\cos{i}\big]
\end{multline}
\begin{multline}\label{eq:ysky}
     Y_\mathrm{rel}(t) = r_\mathrm{tot}\, \big[\cos(\nu(t) +\omega_2)\sin{\Omega}\\
    + \sin(\nu(t)+\omega_2)\cos{\Omega}\cos{i}\big]
\end{multline}

\begin{equation}\label{eq:rtot}
    r_\mathrm{tot} = \frac{a_\mathrm{tot} \, (1 - e^2)}{1 + e \cos \nu}
\end{equation}


RV observations alone provide constraints on $P$, $T_0$, $e$, $\omega_1$, and $K_1$, from which we can also derive the mass function:

\begin{equation}
    f = \frac{{M_2}^3 \sin^3{i}}{(M_1+M_2)^2} = \frac{P {K_1}^3}{2 \pi G}\qty(1 - e^2)^{3/2}
\end{equation}

In the astrometry-only fit, we can derive the values of seven orbital elements: $P$, $T_\mathrm{peri}$, $e$, $\omega_2$, $\Omega$, $i$, and $\atot$.\footnote{The values of the astrometry equations are identical at ($\omega_2, \Omega$) and ($\omega_2 \pm \pi, \Omega \mp \pi$). Therefore, in the astrometry-only fit we restrict $\Omega$ to $[0, \pi)$ and use the results from the RV fit to break the degeneracy between $\Omega$ and $\Omega + \pi$.} From $P$ and $\atot$, Kepler's third law gives us the total mass $M_\mathrm{tot} = M_1 + M_2$. 

With the combination of RVs and relative astrometry observations, we can fully solve for all of the orbital elements of both components, including their individual masses and semi-major axes.


\subsection{RV-only Model}\label{sec:rvonly}

We include RV data from seven instruments, each with their own RV offsets and jitters (see \S\ref{ssec:rvs}). The radial velocity $\mathcal{V}_r$ for each $i$ instrument is

\begin{equation}\label{eq:rvmodel}
    \mathcal{V}_{r,\, i}(t) = K_1 \big[\cos(\nu(t) + \omega_1) + e \cos\omega_1\big] + \gamma_i
\end{equation}

The term $\gamma_i$ represents the RV offset for each individual telescope. For data from the three MINERVA telescopes, we add an additional offset parameter $\Delta_\mathrm{out}$ that represents the shift in the RV baseline that occurred due to a power outage at FLWO on  2021 May 21. The model for the MINERVA data specifically is given by Eq. \ref{eq:rvmodel} prior to 2021 May 21 ($t < 2459347.5$ JD) and 

\begin{equation}
     \mathcal{V}_{r,\, i}(t) = K_1 \big[\cos(\nu(t) + \omega_1) + e \cos\omega_1\big] + \gamma_i + \Delta_\mathrm{out}
\end{equation}

\noindent after that date ($t \geq 2459347.5$ JD).

The log-likelihood is given by

\begin{equation}\label{eq:rvlh}
    \ln{\mathcal{L}_\mathrm{RV}} = \sum_i \log{\mathcal{L}_i}
\end{equation}

\noindent where $\ln \mathcal{L}_i$ is the per-instrument log-likelihood 

\begin{equation}
    \begin{split}
        \ln{\mathcal{L}_i} = - \frac{1}{2}\sum_{j=1}^{N_\mathrm{obs}}\Bigg[\ln&\qty(2 \pi \qty( \sigma_{v,\, j})^2 + \sigma_{\mathrm{jit},\, i}^2) \\
        &+ \frac{(v_j - \mathcal{V}_{r,\, i}(t_j))^2}{\sigma_{v,\, j}^2 + \sigma_{\mathrm{jit},\, i}^2}\Bigg].
    \end{split}
\end{equation}

Here $(t_j, v_j, \sigma_{v,\, j})$ represent the time, velocity, and error of an individual RV measurement and $\sigma_{\mathrm{jit},\, i}$ is the instrumental jitter term. The full set of fitting parameters and priors can be found in Table \ref{tab:all_priors}.

Additionally, the fitting parameter that corresponds to the jitter terms are the variances $\sigma_{\mathrm{jit,\,}i}^2$. The lower limits on these terms are set such that $\sigma_j^2 + \sigma_{\mathrm{jit},\, i}^2 > 0$ for all data from instrument $i$. This means that it is possible for $\sigma_{\mathrm{jit},\, i}^2$ to be zero or negative. We choose these bounds intentionally in order to avoid a Lucy-Sweeney-type bias \citep{Lucy:1971} that systematically disfavors solutions where $\sigma_{\mathrm{jit},\, i} = 0$ and therefore deweights RV measurements with properly or overestimated error bars. This ends up being very important in this case, where we find that the values of $\sigma_{\mathrm{jit},\, i}$ for multiple instruments are consistent with zero (see Section \ref{sec:res}).



\subsection{Astrometry-only Model}\label{sec:astonly}

 We first converted the measured position angle ($\theta_j$) and separation ($\rho_j$) measurements in Table \ref{tab:as} to the relative position of the secondary to the primary in terms of decl. and R.A. ($x_j$ and $y_j$, respectively) using Equation \ref{eq:radec}.
\begin{equation}\label{eq:radec}
    \begin{split}
        x_j &= \rho_j \cos(\theta_j)\\
        y_j &= \rho_j \sin(\theta_j)
    \end{split}
\end{equation}





We model the on-sky orbital position $(X_\mathrm{rel}(t), Y_\mathrm{rel}(t))$ using Equations \ref{eq:xsky} and \ref{eq:ysky}. The model parameters and the priors on these parameters are given in Table \ref{tab:all_priors}. The total log-likelihood of the astrometry model is the sum of the R.A. and decl. components.

\begin{align}
\begin{split}
   \ln{\mathcal{L}_\mathrm{Dec}}    & = -\frac{1}{2} \sum_{j=1}^{N_\mathrm{obs}}\Bigg[\ln\qty(2 \pi\, (S_\mathrm{ast}\,\sigma_{x,\, j})^2) \\
   &\qquad \qquad \qquad + \qty(\frac{x_j - X_\mathrm{rel}(t_j)}{S_\mathrm{ast}\, \sigma_{x,\, j}})^{\!2}\Bigg]
\end{split}\label{eq:ralh}
\\[2ex]
\begin{split}
   \ln{\mathcal{L}_\mathrm{RA}}    & = -\frac{1}{2} \sum_{j=1}^{N_\mathrm{obs}}\Bigg[\ln\qty(2 \pi\, (S_\mathrm{ast}\,\sigma_{y,\, j})^2) \\
   &\qquad \qquad \qquad + \qty(\frac{y_j - Y_\mathrm{rel}(t_j)}{S_\mathrm{ast}\, \sigma_{y,\, j}})^{\!2}\Bigg]
\end{split}\label{eq:declh}
\end{align}

\begin{equation}\label{eq:astlh}
    \ln \mathcal{L}_\mathrm{ast} = \mathbf{\ln} \mathcal{L}_\mathrm{RA} + \mathbf{\ln} \mathcal{L}_\mathrm{Dec}
\end{equation}

\noindent where ($t_j$, $x_j$, $y_j$, $\sigma_{x,\,j}$, $\sigma_{y,\,j}$) are the time, relative position, and error of each observation and $S_\mathrm{ast}$ is a multiplicative error scaling term applied to the error bars on the astrometry data. We note this formulation of the log-likelihood does not account for any covariance between the R.A. and the decl. resulting from the conversion from PA and $\rho$ to R.A. and decl. and therefore this is not included in the fits.


\subsection{Joint RV and Astrometry Model}\label{sec:joint}

When combining RV and relative astrometry data sets, we can rewrite our parameters in terms of the masses of the two stars $M_1$ and $M_2$, giving us a full set of orbital elements $\{ P, T_0, e, \omega_1, \Omega, i, M_1, M_2\}$. These parameters can be input into the RV and astrometry models described in Sections \ref{sec:rvonly} and \S \ref{sec:astonly} using Kepler's third law and Equation \ref{eq:k1} to calculate $\atot$ and $K_1$. The full list of fit parameters and priors is given in Table \ref{tab:all_priors}.



The log-likelihood of the joint model $\ln \mathcal{L}_\mathrm{3D}$ is the sum of the RV log-likelihood (Eq. \ref{eq:rvlh}) and the astrometry log-likelihood (Eq. \ref{eq:astlh}):

\begin{equation}
    \ln \mathcal{L}_\mathrm{3D} = \ln \mathcal{L}_{\mathrm{RV}} + \mathcal{L}_\mathrm{RA} + \mathbf{\ln} \mathcal{L}_\mathrm{Dec}
\end{equation}

\begin{deluxetable*}{LccccC}[!ht]
    \tabletypesize{\footnotesize}
    \tablecaption{List of parameters directly estimated by fitting RV-only, astrometry-only, and joint models to the data set, and the priors on these parameters. Columns 3-5 indicate if the prior is included in the posterior function of the indicated model. Colors are only used to help visually distinguish groups of model parameters. \label{tab:all_priors}}
    \tablehead{
        \colhead{Parameter} & 
        \colhead{Unit} &
        \multicolumn{3}{c}{In model\tablenotemark{\scriptsize{a}}} &
        \colhead{Prior}\\
        \cline{3-5}
        \colhead{} & \colhead{} & \colhead{\scriptsize{RV}} & \colhead{\scriptsize{Astr.}} & \colhead{\scriptsize{Joint}} & \colhead{}
    }
    \startdata
    \multicolumn{6}{l}{System Parameters}\\
    \hline
        \ln{P} & yr & \ry & \ay & \jy & \mathcal{U}(\ln(40), \ln(200))\\
        T_\mathrm{peri} & JD$ - 2450000$ & \ry & \ay & \jy & \mathcal{U}(8222 - P/2, 8222 + P/2)\\
        \sqrt{e} \cos{\omega_1} & & \ry & \ay & \jy & \mathcal{U}(-1, 1)\tablenotemark{\scriptsize{b}}  \\
        \sqrt{e} \sin{\omega_1} & & \ry & \ay & \jy & \mathcal{U}(-1, 1)\tablenotemark{\scriptsize{b}}  \\
        \cos{i}  &   & \rn & \ay & \jy & \mathcal{U}(-1, 1)  \\
        \Omega & \degree & \rn & \ay & \jy & \mathcal{U}(0, 180) \\
        \ln{K_1} & m s$^{-1}$  & \ry & \an & \dy & \mathcal{U}(\ln(500), \ln(1000))\\
        a_\mathrm{tot} & $''$ & \rn & \ay & \dy & \mathcal{U}(0.01, 10)\\
        M_1 &  \msune & \rn & \an & \jy & \mathcal{U}(0, 5) \\
        M_2 &  \msune & \rn & \an & \jy & \mathcal{U}(0, 5) \\[0.25em]
        \hline
        \multicolumn{6}{l}{RV Offset and Jitter Parameters}\\
        \hline
        \gamma_\mathrm{Lick}& m s$^{-1}$  & \ry & \an & \jy & \mathcal{U}(-50, -30) \\ 
        \gamma_\mathrm{HARPS}& m s$^{-1}$ & \ry & \an & \jy & \mathcal{U}(-45, -25)\\ 
        \gamma_\mathrm{HIRES}& m s$^{-1}$ & \ry & \an & \jy & \mathcal{U}(-50, -30)\\ 
        \gamma_\mathrm{APF}& m s$^{-1}$ & \ry & \an & \jy   & \mathcal{U}(-60, -40)\\ 
        \gamma_\mathrm{T1}& m s$^{-1}$ & \ry & \an & \jy    & \mathcal{U}(-65, -45)\\ 
        \gamma_\mathrm{T2}& m s$^{-1}$ & \ry & \an & \jy    & \mathcal{U}(-85, -65)\\ 
        \gamma_\mathrm{T3}& m s$^{-1}$ & \ry & \an & \jy    & \mathcal{U}(-40, -20)\\
        \Delta_\mathrm{out}& m s$^{-1}$ & \ry & \an & \jy   & \mathcal{U}(-30, 0)  \\
        \sigma_{\mathrm{jit},\,i}^2&  m$^2$ s$^{-2}$ & \ry & \an & \jy & \mathcal{U}(-\min(\sigma_{v,\, i})^2, 1000)\tablenotemark{\scriptsize{c}} \\
        \ln{S_\mathrm{ast}} & & \rn & \ay & \jy  & \mathcal{U}(-10, 5)
    \enddata
    \tablenotetext{a}{{\tiny{\faIcon{circle}}}\, = Fitting parameter for which the prior is included in this posterior function of this model, {\tiny{\faIcon[regular]{circle}}}\, = Parameter which can be derived from fitting parameters and for which the prior is not applied, {\small{$\mathbf{\times}$}}\, = Not applicable to this model}
    \vspace{-3pt}
    \tablenotetext{b}{Implicit in these priors is the constraint that $\omega_1 \in (0 \degree, 360\degree)$. We include an additional constraint on eccentricity $e \in [0, 1)$, which is not a necessary condition of the priors.}
    \vspace{-3pt}
    \tablenotetext{\scriptsize{c}}{This line represents the seven instrumental jitter parameters. For each instrument $i$, the prior on the variance $\sigma_{\mathrm{jit}, i}^2$ is a function of the error measurements taken with that instrument $\sigma_{v, i}$ (see \S \ref{sec:rvonly}). }
\end{deluxetable*}

\subsection{Model fitting with PyMC}\label{sec:pymc}

We model the 3D orbit of GJ 105 AC using the No U-Turn Sampler \citep[NUTS;][]{Hoffman2011} implemented in the probabilistic programming library \texttt{PyMC} \citep{pymc:paper, pymc:zenodo}. The NUTS sampler is a variation of the Hamiltonian Monte Carlo (HMC) sampler, a class of  Markov Markov Chain Monte Carlo (MCMC) samplers that uses gradient-informed sampling to reduce inefficiency associated with random walks. NUTS is ideal for sampling models with large numbers of parameters and high-volume parameter spaces.

For each NUTS iteration, the trajectory of the simulated ``Hamiltonian system" is determined by the step size parameter $\epsilon$ and number of ``leapfrog steps" $L$. The step size in each dimension of parameter space can also be scaled by introducing an $n \times n$ ``inverse-mass matrix" $M^{-1}$, where $n$ is the number of parameters in the model. For our model, the step scale for different parameters can be expected to vary by orders of magnitude, and it is important to choose an appropriate scaling matrix.

We developed a method for initializing the inverse-mass matrix prior to the tuning steps by adapting the \texttt{EXOFAST\_GETMCMCSCALE} routine that is described in \citet{Eastman2013}. This routine was developed to determine an appropriate stepping scale for the Differential Evolution Markov Chain method implemented in \texttt{EXOFAST}. Starting at the maximum a posteriori best-fit values, the value of each parameter is varied in small increments until the step size is found that results in $\Delta \chi^2 = 1.$ We use these step scales as the initial values on the diagonal of the inverse-mass matrix. This procedure relies on the incorrect assumption that the errors are Gaussian and uncorrelated, but it provides a much better stepping scale than the default, where $M^{-1} = I$. The code is a direct translation of \texttt{EXOFAST\_MCMCSCALE} from IDL to Python.


\subsection{Reweighting the Astrometry Data}\label{sec:resigma}

For our RV data, we are able to include an additional instrumental jitter term for each instrument because the number of instruments is much smaller than the number of data points. The astrometry data is composed of 13 data points that come from seven different instruments; therefore, including one error scaling term per instrument would create too many degrees of freedom in the fit. However, when we look at the model residuals in comparison to the published error bars on the astrometry data, it is clear that not all of these uncertainties are accurate. We use the following process to derive new, reweighted errors. We then fit our astrometry-only and joint models using both the published errors and the corrected errors to compare the results.

We use a simplified variant of a Leave-One-Out (LOO) cross-validation model evaluation to determine the approximate scale of the errors $\sigma^*_\rho$ from each instrument. For each of the 13 astrometry data points, we remove the $n$th astrometry data point and fit the remaining data to the joint model as described in Sections \ref{sec:joint} and \ref{sec:pymc}. Here we will refer to the best-fit astrometry models with the $k$th point removed as $X_{\mathrm{loo}, k}(t)$ and $Y_{\mathrm{loo}, k}(t)$. From $X$ and $Y$, we can calculate the model separation $\varrho_{\mathrm{loo}, k}(t)$:

\begin{equation}
    \varrho_{\mathrm{loo},k}(t) = \sqrt{X_{\mathrm{loo}, k}(t)^2 + Y_{\mathrm{loo}, k}(t)^2}
\end{equation}

For each data point, we compute the rms deviation (RMSD) of the data as compared to each of the LOO models:



\begin{equation}\label{eq:rmsd}
    r_j = \sqrt{\frac{\sum_{k=1}^{13} (\rho_j - \varrho_{\mathrm{loo}, k}(t_j))^2}{13}}
\end{equation} 

\begin{figure}[!ht]
    \centering
    \includegraphics[width=\columnwidth]{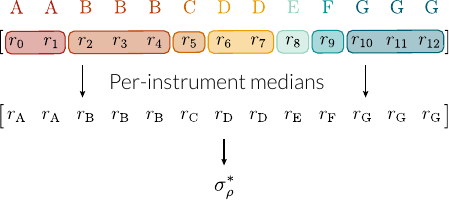}
    \caption{A visual depiction of the process of taking the per-instrument medians of the RMSD to the 13 LOO models (Eq. \ref{eq:rmsd}), as described in \S \ref{sec:resigma}.}
    \label{fig:rmsd_inst}
\end{figure}

For instruments with multiple measurements, the reweighted $\sigma_\rho$ for those data are given by the median RMSD of all data from that instrument (see Figure \ref{fig:rmsd_inst}).

\begin{figure*}[!ht]
    \centering
    \includegraphics{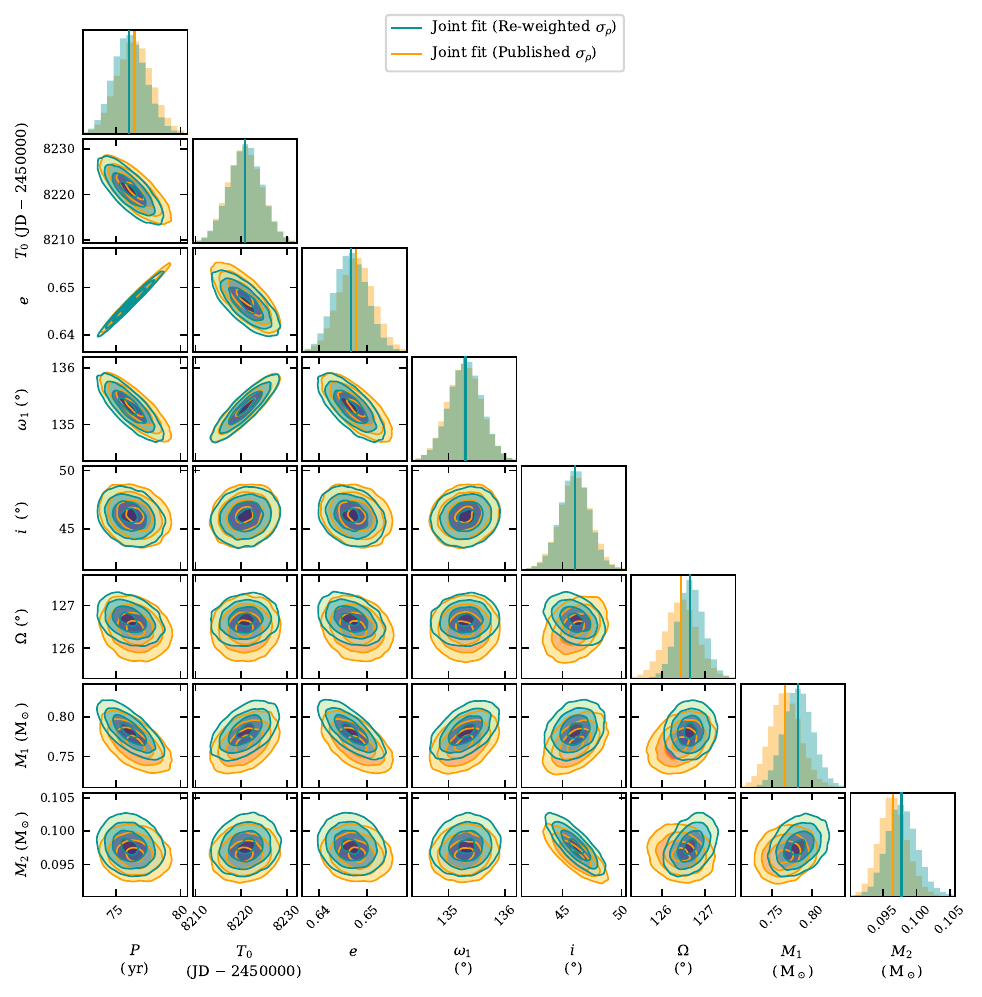}
    \caption{Corner plot showing the posterior distributions of orbital parameters in the joint RV and astrometry model, sampled using the NUTS sampler described in \S \ref{sec:pymc}. The 1D histograms and 2D kernel density contours shown in blue represent the posteriors derived from fitting to the data set with the re-weighted astrometry error bars; those in orange represent the fit to the data set with the published error bars. We can visually determine that the best-fit parameter values do not depend heavily on which set of error bars we apply to the astrometry.}
    \label{fig:corner}
    \vspace{20pt}
\end{figure*}

We run the astrometry-only and joint fits using both the published error bars and our rescaled errors and compare the results in Table \ref{tab:posterior} and Figure \ref{fig:corner}. Overall, the mass measurements of the primary and the companion only change by $2$\% and $1.3$\% respectively, when fitting the joint model to data with the published errors as opposed to the reweighted error bars. The other orbital parameters each change by $\lesssim 0.5$\%.

\section{Results} \label{sec:res}

\begin{figure*}[!t]
    \centering
    \includegraphics{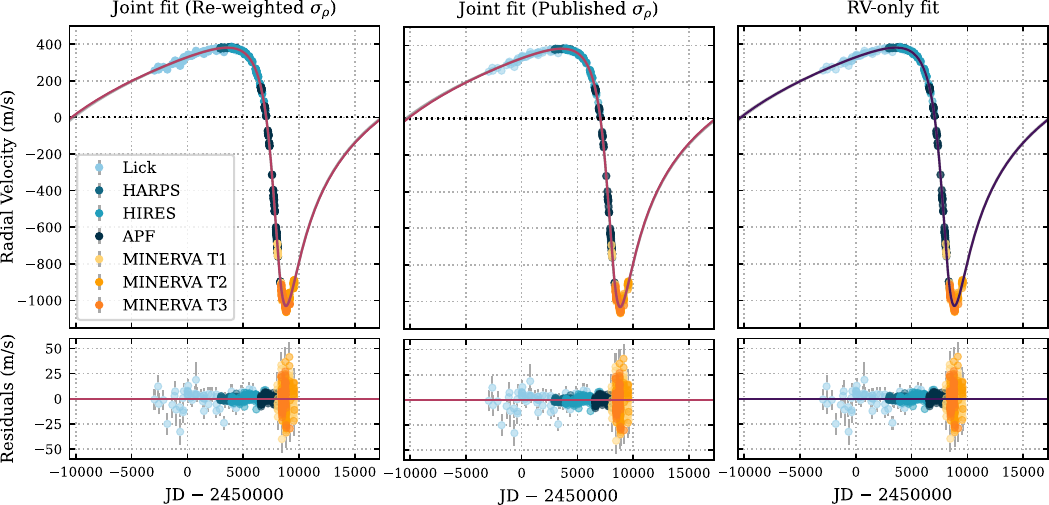}
    \caption{The radial velocity timeseries data, with three different models overplotted: the best-fit joint RV+astrometry model to the data set using the re-weighted values of $\sigma^*_\rho$, the best-fit joint RV+astrometry model to the data set using the published values of $\sigma_\rho$, and the fit to the RV data alone. All three models are plotted on the panels, but they are difficult to distinguish by eye. In the top panels, the model in the plot title is plotted in color and the other two models are plotted in gray. In the bottom panels, the residuals to the highlighted fit are plotted.}
    \label{fig:rvmodel}
    \vspace{10pt}
\end{figure*}

\begin{figure*}[!htb]
    \centering
    \includegraphics[width=\textwidth]{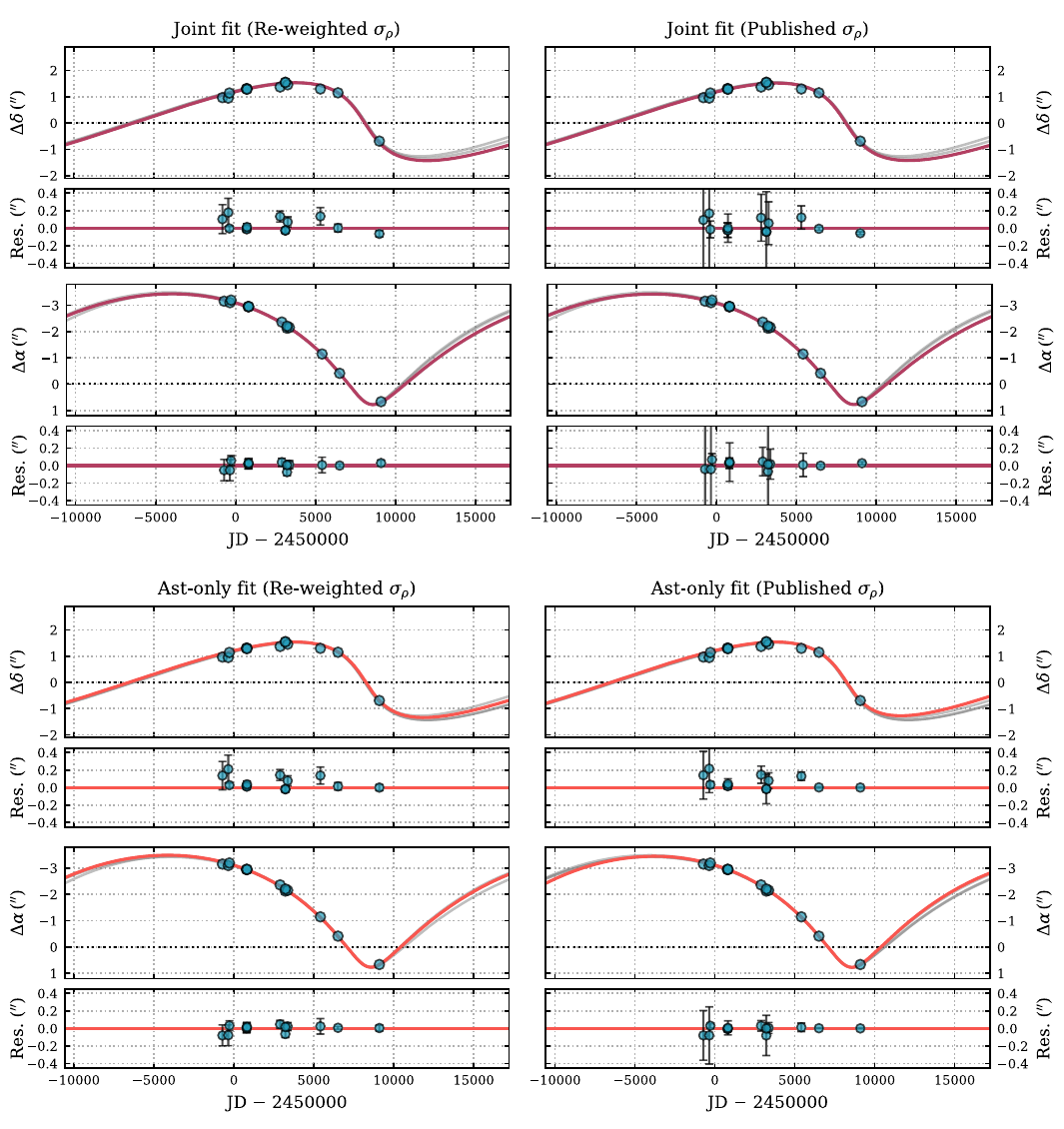}
   \caption{The astrometry timeseries data, with four different models overplotted: the best-fit joint RV+astrometry model 1) to the data set using the re-weighted values of $\sigma^*_\rho$ and 2) to the data set using the published values of $\sigma_\rho$, as well as the best-fit to the astrometry data alone 3) using the re-weighted values of $\sigma^*_\rho$ and 4) using the published values of $\sigma_\rho$. All four best-fit models are shown in each panel, with the model in the title plotted in color and the other three models plotted in gray. For each panel, the top row is a plot of the change in Dec ($\Delta \delta$) data and best-fit models and the second row is a plot of the $\Delta \delta$ residuals to the highlighted model. The third and bottom rows are the same plots as above for the change in RA ($\Delta \alpha$). The error bars on the residual plots reflect the fitted uncertainties multiplied by the value of $S_\mathrm{ast}$ for that model.}
    \label{fig:astmodel}
    \vspace{50pt}
\end{figure*}

\begin{figure*}[p]
    \centering
    \includegraphics[width=0.61\textwidth]{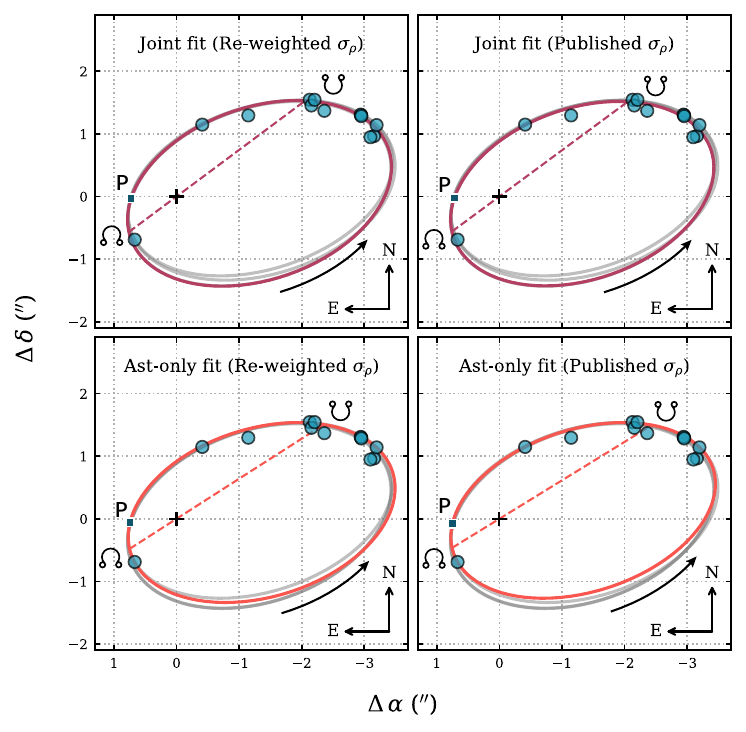}
    \caption{Best-fit 2D orbits of GJ 105 C relative to GJ 105 A. Four models are plotted on each of the four plots: the best-fit joint RV+astrometry model 1) to the data set using the re-weighted values of $\sigma^*_\rho$ and 2) to the data set using the published values of $\sigma_\rho$, as well as the best-fit to the astrometry data alone 3) using the re-weighted values of $\sigma^*_\rho$ and 4) using the published values of $\sigma_\rho$. In each panel, the best-fit model in the title is shown in color and the other 3 models are plotted in gray. The astrometric data are represented by the blue circles. The line of nodes for the highlighted model is depicted by a dashed line} and the ascending node and descending nodes are notated with \includegraphics[width=8pt]{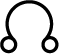} and \includegraphics[width=8pt]{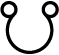} symbols respectively. Periapsis is marked by a ``P".
    \label{fig:ast2dmodel}
\end{figure*}

We determine from the joint RV and astrometry model using the reweighted astrometry errors that GJ 105 C has an orbital period of $76.0 \pm 1.3$ years. We find that all fits yield consistent orbital periods within $\lesssim1.3\sigma$. 
Figure \ref{fig:rvmodel} shows the best-fit joint models and RV-only model in comparison to the RV data, and the residuals to those models and Figure \ref{fig:astmodel} shows the same for the joint models and astrometry-only models with respect to the relative astrometry data. Figure \ref{fig:ast2dmodel} depicts the 2D orbit as seen on sky with the joint models and astrometry-only models overplotted. 
A full list of fit properties is given for the three models and compared in Table \ref{tab:posterior}, with the preferred parameter set highlighted in pink.

We measure the mass of the primary to be $M_A = 0.78 \pm 0.02$ \msun and that of the secondary to be $M_C = 0.098 \pm 0.002$ \msune. Comparing these values to the \texttt{EXOFASTv2} masses in Section \ref{sec:sys}, they are in relatively good agreement ($1.4\sigma$ for both components). This also agrees very well with the mass of GJ 105 C from \citet{Feng2021}  of $M_C = 0.098 \pm 0.009$\msun. 

The dynamically measured masses are much more precise than previous mass measurements, at 2.5\% and 2.3\% respectively, putting them below the $3\%$ precision threshold discussed in Section \ref{sec:intro}. Including the uncertainty introduced by the choice to fit the data using our reweighted error bars or the published error bars, the precision is still $\sim$3\% for both masses. This source of error likely could be mitigated in the future with a small number of additional astrometric observations. This makes GJ 105 C one of the lowest-mass stars with a dynamically measured mass. The suitability of these stars as potential benchmark stars is discussed further in Section \ref{sec:disc}. 

\section{Discussion} \label{sec:disc}

\subsection{Benchmark Stars: When Are Binaries Completely Detached?}\label{sec:iso}

 In Section 1, we introduced the idea that we require better than 3\% precision in stellar mass measurements for benchmark stars. This is due to the precision of the stellar evolution models to which we wish to compare them. By the same logic, when we are modeling stars in multiple-star systems as single stars, we want to ensure that the observed physical properties of the star are not impacted by the presence of its companion star by more than 1-3\%. In this paper, we call stars in binaries that meet this criterion ``effectively single".

Here we consider different ways in which the components of a binary star may impact one another in terms of their stellar evolution and physical properties. In this section we look at three such effects: tidal locking, tidal distortion due to the presence of a close companion, and instellation flux received at the star's surface from its companion. We derive approximate criteria for the required separation between two stars that can be applied to real binaries to estimate the degree to which they can be considered effectively single. While we cannot make this determination precisely for every system, we can apply it to a large sample to estimate the prevalence of effectively single stars within binaries.

\begin{rotatetable*}
\begin{deluxetable*}{lscRlRlRlRl}
    \tabletypesize{\footnotesize}
    \tablecaption{Posterior values of the fitting parameters in the RV-only, astrometry-only and joint fits. The joint model and the astrometry-only model are both fit to two separate data sets: 1) with the re-weighted error bars $\sigma^*_\rho$ on the astrometry (see \S \ref{sec:resigma}) and 2) with the published astrometry errors. Our ``preferred" fit, from which we adopt the orbital parameters is the $\sigma^*_\rho$ joint fit. Columns 3, 5, 7, and 9 show the how well the other four models compare to the joint fit, in units of  $\sigma$.}\label{tab:posterior}
    \tablehead{
        \colhead{Parameter} & 
        \multicolumn{1}{t}{Joint\tablenotemark{\rm \scriptsize{1}}} &
        \colhead{} & 
        \multicolumn{2}{c}{Joint\tablenotemark{\rm \scriptsize{2}}}&
        \multicolumn{2}{c}{RV-only} &
        \multicolumn{2}{c}{Ast-only\tablenotemark{\rm \scriptsize{1}}} &
        \multicolumn{2}{c}{Ast-only\tablenotemark{\rm \scriptsize{2}}}\\
        \cline{2-2}\cline{4-11}
        \colhead{} & 
        \multicolumn{1}{t}{(Preferred fit)} & 
        \colhead{} & 
        \colhead{Value} & \colhead{($\sigma$)} &
        \colhead{Value} & \colhead{($\sigma$)} &
        \colhead{Value} & \colhead{($\sigma$)} &
        \colhead{Value} & \colhead{($\sigma$)}}
    \decimals
    \startdata        
\multicolumn{11}{l}{System Parameters}\\
\cline{1-2} \cline{4-11}
$P$ (yr) & 76.0\ \substack{+1.3 \\ -1.2} &  & 76.4\ \substack{+1.4 \\ -1.3} & 0.2 & 75.5\ \substack{+1.4 \\ -1.3} & 0.3 & 74\ \substack{+7 \\ -6} & 0.3 & 72 \pm 3 & 1.3\\ 
$T_\mathrm{peri}$ ($\mathrm{JD} - 2450000$) & 8221 \pm 4 &  & 8221 \pm 4 & 0.03 & 8222 \pm 4 & 0.2 & 8330 \pm 60 & 1.7 & 8335 \pm 16 & 6.6\\ 
$e$ & 0.647 \pm 0.003 &  & 0.648 \pm 0.004 & 0.2 & 0.645 \pm 0.004 & 0.3 & 0.645\ \substack{+0.020 \\ -0.017} & 0.06 & 0.641 \pm 0.008 & 0.7\\ 
$\omega_1$ ($\degree$) & 135.3 \pm 0.3 &  & 135.3 \pm 0.3 & 0.04 & 135.4 \pm 0.3 & 0.2 & 143\ \substack{+6 \\ -7} & 1.2 & 145.2 \pm 1.7 & 5.6\\ 
$i$ ($\degree$) & 46.0\ \substack{+1.2 \\ -1.3} &  & 46.1 \pm 1.3 & 0.03 & \multicolumn{1}{c}{--} & -- & 44 \pm 3 & 0.5 & 45.9 \pm 0.9 & 0.09\\ 
$\Omega$ ($\degree$) & 126.7 \pm 0.3 &  & 126.4 \pm 0.4 & 0.4 & \multicolumn{1}{c}{--} & -- & 122 \pm 3 & 1.2 & 121.5 \pm 0.7 & 6.6\\ 
$K_1$ (m/s) & 706.1 \pm 1.0 &  & 706.2 \pm 1.0 & 0.05 & 706.1 \pm 1.0 & 0.003 & \multicolumn{1}{c}{--} & -- & \multicolumn{1}{c}{--} & --\\ 
$\atot$ ($''$) & 2.380 \pm 0.019 &  & 2.37 \pm 0.02 & 0.2 & \multicolumn{1}{c}{--} & -- & 2.31\ \substack{+0.10 \\ -0.07} & 0.7 & 2.30 \pm 0.03 & 2.1\\ 
$M_1$ (\msune) & 0.782\ \substack{+0.019 \\ -0.018} &  & 0.766 \pm 0.019 & 0.6 & \multicolumn{1}{c}{--} & -- & \multicolumn{1}{c}{--} & -- & \multicolumn{1}{c}{--} & --\\ 
$M_2$ (\msune) & 0.098 \pm 0.002 &  & 0.096 \pm 0.002 & 0.4 & \multicolumn{1}{c}{--} & -- & \multicolumn{1}{c}{--} & -- & \multicolumn{1}{c}{--} & --\\ 
\cline{1-2} \cline{4-11}
\multicolumn{11}{l}{RV Offset Parameters}\\
\cline{1-2} \cline{4-11}
$\gamma_\mathrm{Lick}$ (m s$^{-1}$) & -39.4 \pm 1.5 &  & -39.2 \pm 1.5 & 0.005 & -39.4 \pm 1.5 & 0.0012 & \multicolumn{1}{c}{--} & -- & \multicolumn{1}{c}{--} & --\\ 
$\gamma_\mathrm{HARPS}$ (m s$^{-1}$) & -33.2 \pm 1.5 &  & -32.7 \pm 1.5 & 0.015 & -33.5 \pm 1.5 & 0.011 & \multicolumn{1}{c}{--} & -- & \multicolumn{1}{c}{--} & --\\ 
$\gamma_\mathrm{HIRES}$ (m s$^{-1}$) & -37.6 \pm 1.7 &  & -37.0 \pm 1.7 & 0.015 & -37.9 \pm 1.7 & 0.010 & \multicolumn{1}{c}{--} & -- & \multicolumn{1}{c}{--} & --\\ 
$\gamma_\mathrm{APF}$ (m s$^{-1}$) & -45.5 \pm 1.7 &  & -45.0\ \substack{+1.7 \\ -1.6} & 0.010 & -45.8 \pm 1.6 & 0.007 & \multicolumn{1}{c}{--} & -- & \multicolumn{1}{c}{--} & --\\ 
$\gamma_\mathrm{T1}$ (m s$^{-1}$) & -52.6 \pm 1.9 &  & -52.0 \pm 1.9 & 0.011 & -53.0 \pm 1.9 & 0.007 & \multicolumn{1}{c}{--} & -- & \multicolumn{1}{c}{--} & --\\ 
$\gamma_\mathrm{T2}$ (m s$^{-1}$) & -72 \pm 2 &  & -71 \pm 2 & 0.008 & -72 \pm 2 & 0.005 & \multicolumn{1}{c}{--} & -- & \multicolumn{1}{c}{--} & --\\ 
$\gamma_\mathrm{T3}$ (m s$^{-1}$) & -30 \pm 2 &  & -30 \pm 2 & 0.019 & -30 \pm 2 & 0.010 & \multicolumn{1}{c}{--} & -- & \multicolumn{1}{c}{--} & --\\ 
$\Delta_\mathrm{out}$ (m s$^{-1}$) & -20 \pm 4 &  & -20 \pm 4 & 0.03 & -21 \pm 4 & 0.1 & \multicolumn{1}{c}{--} & -- & \multicolumn{1}{c}{--} & --\\ 
\cline{1-2} \cline{4-11}
\multicolumn{11}{l}{Jitter Parameters}\\
\cline{1-2} \cline{4-11}
$\sigma^2_\mathrm{Lick}$ (m$^2$ s$^{-2}$) & -19\ \substack{+12 \\ -9} &  & -20\ \substack{+12 \\ -9} & 0.03 & -18\ \substack{+13 \\ -10} & 0.05 & \multicolumn{1}{c}{--} & -- & \multicolumn{1}{c}{--} & --\\ 
$\sigma^2_\mathrm{HARPS}$ (m$^2$ s$^{-2}$) & -0.4\ \substack{+0.8 \\ -0.6} &  & -0.4\ \substack{+0.8 \\ -0.6} & 0.02 & -0.4\ \substack{+0.8 \\ -0.6} & 0.017 & \multicolumn{1}{c}{--} & -- & \multicolumn{1}{c}{--} & --\\ 
$\sigma^2_\mathrm{HIRES}$ (m$^2$ s$^{-2}$) & -0.5\ \substack{+1.6 \\ -1.3} &  & -0.4\ \substack{+1.5 \\ -1.2} & 0.03 & -0.5\ \substack{+1.5 \\ -1.3} & 0.05 & \multicolumn{1}{c}{--} & -- & \multicolumn{1}{c}{--} & --\\ 
$\sigma^2_\mathrm{APF}$ (m$^2$ s$^{-2}$) & 7.9\ \substack{+1.2 \\ -1.0} &  & 7.9\ \substack{+1.2 \\ -1.1} & 0.0005 & 7.9\ \substack{+1.2 \\ -1.1} & 0.004 & \multicolumn{1}{c}{--} & -- & \multicolumn{1}{c}{--} & --\\ 
$\sigma^2_\mathrm{T1}$ (m$^2$ s$^{-2}$) & 170\ \substack{+17 \\ -15} &  & 170\ \substack{+18 \\ -15} & 0.0005 & 169\ \substack{+18 \\ -16} & 0.0016 & \multicolumn{1}{c}{--} & -- & \multicolumn{1}{c}{--} & --\\ 
$\sigma^2_\mathrm{T2}$ (m$^2$ s$^{-2}$) & 210 \pm 30 &  & 210 \pm 30 & 0.0007 & 210 \pm 30 & 0.002 & \multicolumn{1}{c}{--} & -- & \multicolumn{1}{c}{--} & --\\ 
$\sigma^2_\mathrm{T3}$ (m$^2$ s$^{-2}$) & 130\ \substack{+30 \\ -20} &  & 130\ \substack{+30 \\ -20} & 0.0011 & 140\ \substack{+30 \\ -20} & 0.0011 & \multicolumn{1}{c}{--} & -- & \multicolumn{1}{c}{--} & --\\ 
$\ln S_\mathrm{ast}$ & 0.37\ \substack{+0.16 \\ -0.14} &  & 1.85\ \substack{+0.16 \\ -0.14} & 6.8 & \multicolumn{1}{c}{--} & -- & 0.34\ \substack{+0.17 \\ -0.16} & 0.1 & 0.84\ \substack{+0.18 \\ -0.16} & 2.0 
\\[0.25em]
\enddata
\tablenotetext{1}{Fit uses the re-weighted astrometry errors, $\sigma^*_\rho$}
\tablenotetext{2}{Fit uses the published astrometry errors}
\end{deluxetable*}
\end{rotatetable*}
 
\begin{figure*}[p]
    \centering
    \includegraphics[width=0.6\textwidth]{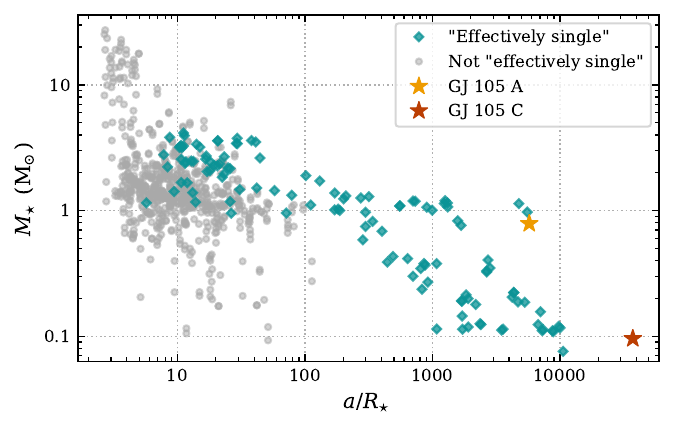}
    \caption{Plot of the stellar mass $M_\star$ vs.semi-major axis divided by the stellar radius $(a/R_\star)$ for the same stars in of the sample of \nstars precise-mass stars described in \S \ref{sec:intro}. The points in this plot are color-coded to demonstrate how many of these stars are ``effectively single", based on the criteria described in \S \ref{sec:iso}. Of the \nstars stars included in this sample, only 122 (15\%) are estimated to be effectively single. GJ 105 A and C are also plotted.}
    \label{fig:iso}
\end{figure*}

With $\atot = 3740\,$\rsune, the components of GJ 105 AC have a very large physical separation that makes them extremely unlikely to interact with one another. In order to consider this system in the context of a broader population of similar binaries, we use the above-described criteria to compare it to the well-studied sample from DEBCat and other sources (see Section \ref{sec:intro}). All of these stars have masses measured to 3\% precision or better. The stars that pass all three tests are shown in Figure \ref{fig:iso}.

\subsubsection{Tidal Locking}\label{sec:tidallock}

 \citet{Fleming2019} used simulations to model the effects of tidal interactions on the rotation of low-mass binary stars. They find that most systems with $P \lesssim 20$ days are tidally locked, for a range of values of tidal quality factor $Q$ and  stellar convective turnover timescale $\tau$. Additionally, they find that binaries up to at least $P = 80$ days are likely to be ``tidally interacting" (i.e. their rotation period is within 10\% of the equilibrium period $P_{\rm eq}$.) 

From Kepler's third law we can derive the semi-major axis at which a system will be tidally isolated as a function of the total mass $M_{\rm tot}$. 

\begin{equation}
    \atot \gtrsim \qty(\frac{G M_{\rm tot}}{4 \pi^2})^{1/3} \times (80\, \text{days})^{2/3}
\end{equation}

\begin{figure}[!ht]
    \centering
    \includegraphics[width=\columnwidth]{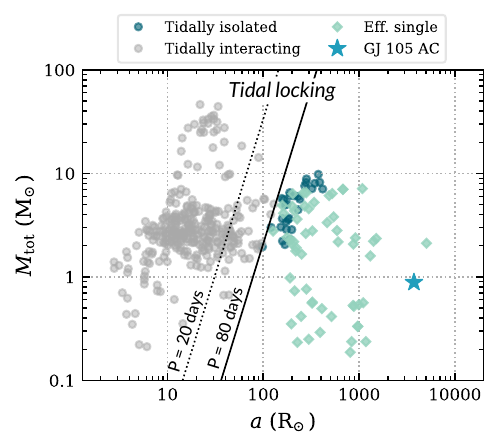}
    \caption{We plot the total mass $M_{\rm tot} = M_1 + M_2$ vs. semi-major axis in of the sample of \nsys precise-mass binary/triple systems shown in Figure \ref{fig:blend}, color-coded based on tidal isolation criteria for ``effectively single" described in \S \ref{sec:tidallock}. The gray points represent systems that are tidally ``interacting" or tidally locked, and the blue, circular points represent systems that are tidally isolated, i.e. those for which $P > 80\,$days \citep{Fleming2019}. The lighter blue, diamond-shaped points indicate the systems where both stars are effectively single by all three criteria. The dotted and solid black lines represent where we expect stars to be tidally locked ($P < 20$ days) and interacting ($P < 80$ days) respectively.}
    \label{fig:single1}
\end{figure}

By this metric, it is difficult to analytically set a lower limit in terms of $a/R_\star$. In general, binary components that are separated by at least $100 R_\star$ appear to be tidally isolated, and all of the low-mass stars in our sample with $a/R_\star < 100$ are isolated. This suggests that for low-mass stars a reasonable criterion of ``effectively single" is $a/R_\star \gtrsim 100$. Alternatively, $a > 100 R_\sun$ could be a more general criterion (see Figure \ref{fig:single1}).

\subsubsection{Tidal Distortion}\label{sec:tidaldist}

In the presence of a close companion, the tidal force may cause a star to become elongated along one axis and become prolate. Here we define a coordinate system centered on the affected star. A prolate star has  rotational symmetry about the major axis, so we can simplify this system to two dimensions: the $x$-axis pointed toward the center of mass of the binary and the $y$-axis perpendicular to that. The change in radius $\delta r$ at a point on the surface of the star due to a perturbation to the gravitational potential around a star is given by

\begin{equation}
    \delta r = \frac{m}{M_\star} \qty(\frac{R_\star}{a})^3\, \frac{R_\star}{2}\, \qty(3\cos^2 \psi - 1)
\end{equation}

\noindent where $\psi$ is the elevation angle with respect to the $x$-axis, $M_\star$ and $R_\star$ are the mass and radius of the body, $m$ is the mass of the perturber, and $a$ is the separation \citep{Bradt2008}.

We define 
$r_x$ and $r_y$ as the distances to the body's surface at $\psi=0\degree$ and $\psi = 90\degree$, respectively. We can measure the distortion of a star due to its companion as the ellipticity, or flattening, $\varepsilon$:

\begin{equation}
    \varepsilon = 1 - \frac{r_y}{r_x}
\end{equation}

\noindent where 

\begin{equation}
    \begin{split}
        r_x &= R_\star\qty[1 + \frac{m}{M_\star} \qty(\frac{R_\star}{a})^3]\\
        r_y &= R_\star\qty[1 - \frac{1}{2}\frac{m}{M_\star} \qty(\frac{R_\star}{a})^3]
    \end{split}
\end{equation}

Simplifying these equations, we find that 

\begin{equation}
    \varepsilon \approx \frac{3}{2} \frac{m}{M_\star} \qty(\frac{R_\star}{a})^3
\end{equation}

The flux that we measure is directly proportional to the cross-sectional area of the star as viewed by an observer, $A = \pi R_\star^2$. For a nonspherical star, that cross-sectional area changes based on the viewing angle. The maximum area is $A_\mathrm{max} = \pi r_y r_x$ and the minimum area is $A_\mathrm{min} = \pi r_y^2$. Assuming that the volume of the star remains constant, we find that 

\begin{equation}
    \delta A = A_\mathrm{max} - A_\mathrm{min} 
\end{equation}

\begin{equation}
    \delta A = \varepsilon \cdot \pi R_\star^2
\end{equation}

\noindent Therefore, 1\% ellipticity corresponds to a 1\% variation in stellar flux. 

\begin{figure}[!ht]
    \centering
    \includegraphics[width=\columnwidth]{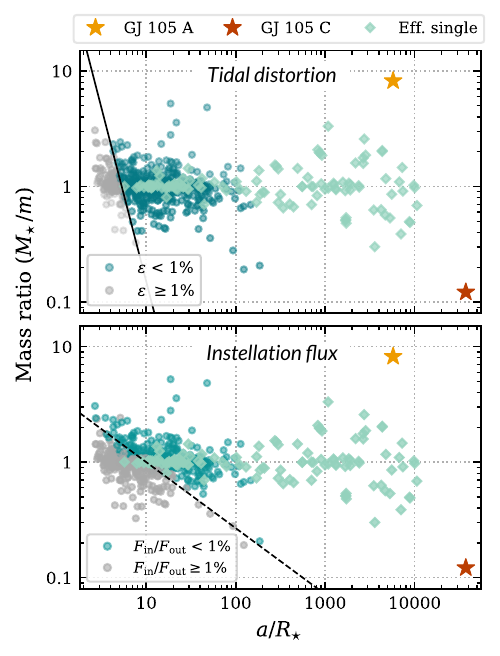}
    \caption{\textit{Top:} We plot the ratio between a star's mass $M_\star$ and the mass of its companion $m$ vs. semi-major axis divided by the stellar radius ($a/R_\star$) for the same set of binary stars in of the sample of \nstars precise-mass stars. The points are color-coded based on tidal distortion criteria for ``effectively single" described in \S \ref{sec:tidaldist}. The blue, circular points represent stars that are effectively single by this criteria (i.e. $\varepsilon <$ 1\%) and the gray points are stars that do not meet this criteria. The black line represents the estimated $\varepsilon = 1$\% boundary. The lighter blue, diamond-shaped points indicate the stars are effectively single by all three criteria.
    \textit{Bottom:} The same as above for the instellation flux criteria described in in \S \ref{sec:instell}. Here, the black dashed line is a rough boundary line, above which we expect a star to receive less than $0.01  F_\star$ at its surface as instellation flux from its companion, assuming a mass-luminosity relationship of $L \propto M^{3.5}$.}
    \label{fig:dist_flux}
\end{figure}

When the masses of the two stars are roughly equivalent, $\varepsilon = 0.01$ when $a/R_1 = a/R_2 \approx 5$. For a system like GJ 105 AC, where $q \equiv M_2/M_1 \approx 1/8$, the ellipicity of the primary $\varepsilon_1$ would be 0.01 when $a/R_1 \approx 2.7$ and the ellipticity of the secondary would be $\varepsilon_2 = 0.01$ when $a/R_2 \approx 10$ (see Figure \ref{fig:dist_flux}).

\subsubsection{Instellation Flux}\label{sec:instell}

Irradiated stars have been shown to evolve differently than their non-irradiated counterparts. This has been modeled in the extreme cases of stars in close proximity to a quasar \citep{Tout1989} and stars undergoing radiative feedback from a Dyson sphere \citep{Huston2022}. The ratio of instellation flux to surface flux is

\begin{equation}
    \frac{F_\mathrm{in}}{F_\mathrm{out}} = \frac{L_2}{L_1} \qty(\frac{R_1}{a})^2.
\end{equation}

For stars whose companion has a luminosity $L_2 = L_1$, the instellation flux at the surface is 1\% of the outgoing flux when $a/R_1 = 10$. As the flux ratio between the primary and secondary increases, the distance at which the primary is suitable decreases, but the companion is more likely to be significantly irradiated (see Figure \ref{fig:dist_flux}).

\vspace{-5pt}

\subsection{GJ 105 A as a Benchmark Star}

With this measurement of the dynamical masses, GJ 105 A and C are now two of the only effectively single stars for which we may be able to obtain clean spectra with such precise masses (see Figure \ref{fig:blend}). The very large ($\sim$400:1) luminosity ratio of the binary makes it very easy to obtain clean spectra of GJ 105 A. Even in spite of this, it may be possible to obtain spectra of GJ 105 C with less than 1\% dilution from the primary in the future. GJ 105 AC has an on-sky separation greater than any binary system other than $\alpha$ Centauri. From the orbital components measured in Section \ref{sec:res}, we calculate the average on-sky separation between GJ 105 A and GJ 105 C to be $2\farcs67$ and the maximum on-sky separation to be $3\farcs51$. The stars recently passed through periapsis, so their separation will be increasing over the next $\sim\,$30 years.

\vspace{-5pt}

\subsection{Future Work: Asteroseismology}

As a bright, nearby star, GJ 105 A may be a possible candidate for asteroseismology follow-up to determine the age of the system and independently confirm the star's mass. We consider the feasibility of detecting $p$-mode oscillations using photometry from the \textit{Transiting Exoplanet Survey Satellite} \citep[TESS;][]{Ricker2015} data or using extremely precise RVs (EPRVs).

We estimate the oscillation frequency $\nu_\mathrm{max}$ and the large oscillation spacing $\Delta \nu$ using the following scaling relations  \citep{Brown1991, Kjeldsen1995}:

\begin{align}
    \nu_\mathrm{max} &= \nu_{\mathrm{max}, \odot} \qty(\frac{g}{g_\odot})\qty(\frac{T_\mathrm{eff}}{T_{\mathrm{eff} ,\odot}})^{-0.5}\\
    \Delta \nu &= \Delta \nu_\odot \qty(\frac{\rho}{\rho_\odot})^{1/2}
\end{align}

Assuming $\nu_{\mathrm{max}, \odot} = 3090\, \mu$Hz and $\Delta \nu_\odot = 135.1 \, \mu$Hz \citep{Campante2016}, we find that for GJ 105 A $\nu_\mathrm{max} = 4745\, \mu$Hz and $\Delta \nu = 232.0\, \mu$Hz.

The oscillation amplitude of the signal can also be estimated for both TESS and for EPRV data. We use scaling relationship for TESS from \citet{Huber2011}:

\begin{equation}
    A_\mathrm{phot} = A_{\mathrm{phot,}\odot} \qty(\frac{L}{L_\odot})^s \qty(\frac{M_\odot}{M})^t \qty(\frac{T_{\mathrm{eff}, \odot}}{T_\mathrm{eff}})^{r-1} \frac{1}{c_K(T_\mathrm{eff})}
\end{equation}

\noindent where $s=0.838$, $t=1.32$, $r=2$, and 

\begin{equation}
    c_K = \qty(\frac{T_\mathrm{eff}}{5934\, \text{K}})^{0.8}
\end{equation}

\begin{figure}[!tb]
    \centering
    \includegraphics[width=\columnwidth]{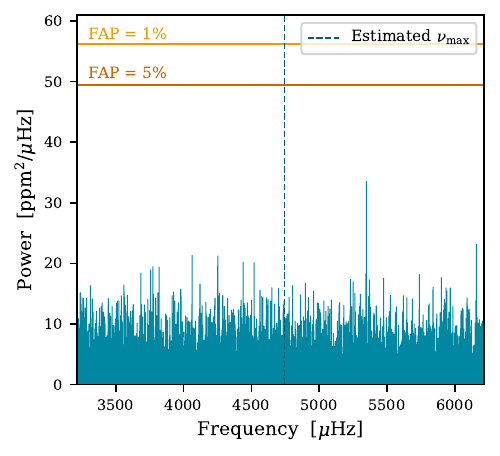}
    \caption{Lomb-scargle periodogram of GJ 105 A from \textit{TESS} sectors 70 and 71. The dashed vertical line is at the predicted frequency of p-mode oscillations $\nu = 4745\, \mu$Hz. We plot the expected power of signals with 1\% and 5\% false alarm probability (FAP) in yellow and orange respectively. There is no signal evident at this predicted frequency  or in the range of frequencies that we explored.}
    \label{fig:pgram}
\end{figure}

Using $A_{\mathrm{phot,}\odot} = 2.125\,$ ppm \citep{Campante2016}, we find that $A_\mathrm{phot,\, pred} = 1.31\,$ppm, which is very unlikely to be observable in TESS photometry. GJ 105 A has been observed in six TES sectors, two of which (sectors 70 and 71) include 20 second cadence observations. We used \texttt{lightkurve} to download the Science Processing Operations Center \citep{Jenkins2016} light curve and generate a periodogram around the predicted oscillation frequency. We found no signal at the expected frequency that is detectable by eye 
(see Figure \ref{fig:pgram}).

We also calculated the expected oscillation amplitude in EPRV data using the scaling relation for NEID in \citet{Gupta2022}:

\begin{equation}
    A_\mathrm{max,\, RV} = A_{\mathrm{max,\, RV,} \odot} \qty(\frac{L_\star}{L_\odot}) \qty(\frac{M_\odot}{M_\star})
\end{equation}

The relationship between $A_\mathrm{max,\, RV}$ and $A_\mathrm{RV}$ is

\begin{equation}
    A_\mathrm{RV} = \sqrt{A_\mathrm{max,\, RV}\, \qty(\frac{\Delta \nu}{c})}
\end{equation}

Adopting $A_{\mathrm{max,\, RV,} \odot} = 0.19\ \mathrm{m}^2\, \mathrm{s}^{-2} \, \mu\mathrm{Hz}^{-1}$ from \citet{Chaplin2019} and $c = 4.09$ from \citet{Kjeldsen2008}, we find that $A_\mathrm{RV,\, pred} = 1.86$ m/s, making GJ 105 A a viable candidate for RV asteroseismology. 

\section{Conclusion} \label{sec:conclusion}

In this paper, we present an updated solution to the orbit of GJ 105 AC with an orbital period of $P = 76.0 \pm 1.3$ years. With the combination of RV and relative astrometry data, we are able to derive dynamical masses of $M_1 = 0.78 \pm 0.02$ and $M_2 = 0.098 \pm 0.002$. The precision of these mass measurements puts GJ 105 AC in a rare class of stars that can be used as points of comparison to models of stellar structures and evolution.

We argue further that the wide on-sky separation and large physical separation from its companion make GJ 105 A an ideal benchmark star and a prime candidate for more detailed follow-up. We were unable to detect $p$-mode oscillations in the two existing sectors of 20 s cadence TESS data, but asteroseismology may be possible in the future using EPRVs. This would be an especially interesting prospect because constraining the age of the primary would also give us an age for the secondary, which would be nearly impossible to do otherwise for a late M star.

\section*{Acknowledgements}

    The authors would like to thank Eric Mamajek for helpful feedback on this paper. 
    
    Work by C.M.D. was supported by the National Aeronautics and Space Administration (XRP 80NSSC22K0233.)

     MINERVA is a collaboration among the Harvard-Smithsonian Center for Astrophysics, the Pennsylvania State University, the University of Montana, the University of Southern Queensland, University of Pennsylvania, George Mason University, and the University of New South Wales. It is made possible by generous contributions from its collaborating institutions and Mt. Cuba Astronomical Foundation, the David \& Lucile Packard Foundation, the National Aeronautics and Space Administration (EPSCOR grant NNX13AM97A, XRP 80NSSC22K0233), the Australian Research Council (LIEF grant LE140100050), and the National Science Foundation (grants 1516242, 1608203, and 2007811). Any opinions, findings, and conclusions or recommendations expressed are those of the author and do not necessarily reflect the views of the National Science Foundation.

    This paper includes data collected by the TESS mission. Funding for the TESS mission is provided by the NASA's Science Mission Directorate. The TESS data referenced in this work can be found in MAST: \dataset[10.17909/t9-st5g-3177]{http://dx.doi.org/10.17909/t9-st5g-3177}.
    
    This research has made use of the Keck Observatory Archive (KOA), which is operated by the W. M. Keck Observatory and the NASA Exoplanet Science Institute (NExScI), under contract with the National Aeronautics and Space Administration. The Keck observations used in this paper were taken as part of the following programs: C17N2 (PI: S. Kulkarni), N096N2L (PI: T. Barclay)and, N179 (PI: E. Gonzales).

    This research has made use of the SVO Filter Profile Service ``Carlos Rodrigo", funded by MCIN/AEI/10.13039/501100011033/ through grant PID2020-112949GB-I00.

    This publication makes use of data products from the Two Micron All Sky Survey, which is a joint project of the University of Massachusetts and the Infrared Processing and Analysis Center/California Institute of Technology, funded by the National Aeronautics and Space Administration and the National Science Foundation.

    This publication makes use of data products from the Wide-field Infrared Survey Explorer, which is a joint project of the University of California, Los Angeles, and the Jet Propulsion Laboratory/California Institute of Technology, and NEOWISE, which is a project of the Jet Propulsion Laboratory/California Institute of Technology. WISE and NEOWISE are funded by the National Aeronautics and Space Administration.

    This work has made use of data from the European Space Agency (ESA) mission {\it Gaia} (\url{https://www.cosmos.esa.int/gaia}), processed by the Gaia Data Processing and Analysis Consortium (DPAC, \url{https://www.cosmos.esa.int/web/gaia/dpac/consortium}). Funding for the DPAC has been provided by national institutions, in particular the institutions participating in the Gaia Multilateral Agreement.

    The Center for Exoplanets and Habitable Worlds is supported by the Pennsylvania State University and the Eberly College of Science.

    MINERVA is located on the ancestral lands of the Tohono O'odham and Hia-Ced O'odham nations; the Ak-Chin Indian Community, and Hohokam people. 

\facilities{MINERVA, TESS, APF, ESO:3.6m (HARPS), Keck:I (HIRES), Shane (Hamilton spectrograph, IRCAL), AEOS, Hale (PHARO), HST (NICMOS, WFPC2), PO:1.5m (AOC).}

\software{arviz \citep{arviz}, astropy \citep{astropy:2013, astropy:2018}, astroquery \citep{astroquery}, EXOFASTv2 \citep{Eastman2019}, \textsf{exoplanet} \citep{exoplanet:joss, exoplanet:zenodo}, matplotlib, \citep{matplotlib}, lightkurve \citep{lightkurve}, numpy \citep{numpy},  pandas \citep{pandas}, pymc \citep{pymc:paper, pymc:zenodo}, pytensor \citep{pytensor:zenodo}, scipy \citep{scipy}.}

\eject

\bibliographystyle{aasjournal}
\bibliography{references,software}

\end{document}